\documentclass[lettersize,journal]{IEEEtran}
\usepackage{amsmath,amsfonts}
\usepackage{algorithm,algpseudocode}
\usepackage{array}
\usepackage[caption=false,font=normalsize,labelfont=sf,textfont=sf]{subfig}
\usepackage{textcomp}
\usepackage{stfloats}
\usepackage{url}
\usepackage{verbatim}
\usepackage{cite}
\usepackage{xcolor}
\usepackage{makecell}
\usepackage{multirow}
\usepackage{cite}
\usepackage{float}
\usepackage{todonotes}
\usepackage{enumerate, enumitem}
\usepackage{graphicx,tabularx}
\usepackage{diagbox,threeparttable,booktabs}

\newcommand{\ie}{\textit{i.e.}, }
\newcommand{\eg}{\textit{e.g.}, }

\newcolumntype{Y}{>{\centering\arraybackslash}X}

\begin{document}

\title{Explainable fMRI-based Brain Decoding via Spatial Temporal-pyramid Graph Convolutional Network}

\author{Ziyuan Ye,~\IEEEmembership{Student Member, IEEE,} Youzhi Qu, Zhichao Liang, Mo Wang, and Quanying Liu*

\thanks{Z. Ye, Y. Qu, Z. Liang, M. Wang, and Q. Liu are with Shenzhen Key Laboratory of Smart Healthcare Engineering, Department of Biomedical Engineering, Southern University of Science and Technology, No. 1088, Xueyuan Rd., Xili, Nanshan District, Shenzhen, Guangdong, 518055, P. R. China.}
\thanks{This work was funded in part by the National Natural Science Foundation of China (62001205), National Key Research and Development Program of China (2021YFF1200800), Guangdong Natural Science Foundation Joint Fund (2019A1515111038), Shenzhen Science and Technology Innovation Committee (20200925155957004, KCXFZ2020122117340001), Shenzhen-Hong Kong-Macao Science and Technology Innovation Project (SGDX2020110309280100), Shenzhen Key Laboratory of Smart Healthcare Engineering (ZDSYS20200811144003009).}
\thanks{* Corresponding author: Quanying Liu (E-mail: {liuqy@sustech.edu.cn}).}
}

\markboth{Ye \MakeLowercase{\textit{et al.}}: STpGCN for interpretable brain decoding}%
{IEEE TNNLS}%

\maketitle

\begin{abstract}
Brain decoding, aiming to identify the brain states using neural activity, is important for cognitive neuroscience and neural engineering. 
However, existing machine learning methods for fMRI-based brain decoding either suffer from low classification performance or poor explainability. 
Here, we address this issue by proposing a biologically inspired architecture, Spatial Temporal-pyramid Graph Convolutional Network (STpGCN), to capture the spatial-temporal graph representation of functional brain activities.
By designing multi-scale spatial-temporal pathways and bottom-up pathways that mimic the information process and temporal integration in the brain, STpGCN is capable of explicitly utilizing the multi-scale temporal dependency of brain activities via graph, thereby achieving high brain decoding performance.
Additionally, we propose a sensitivity analysis method called BrainNetX to better explain the decoding results by automatically annotating task-related brain regions from the brain-network standpoint. 
We conduct extensive experiments on fMRI data under 23 cognitive tasks from Human Connectome Project (HCP) S1200. 
The results show that STpGCN significantly improves brain decoding performance compared to competing baseline models; BrainNetX successfully annotates task-relevant brain regions. Post hoc analysis based on these regions further validates that the hierarchical structure in STpGCN significantly contributes to the explainability, robustness and generalization of the model.
Our methods not only provide insights into information representation in the brain under multiple cognitive tasks but also indicate a bright future for fMRI-based brain decoding.
\end{abstract}


\section{Introduction}
\IEEEPARstart{H}{ow} the brain integrates and segregates the information during a cognitive process remains an opening explorative question in neuroscience~\cite{deco2015rethinking}. With the development of neuroimaging techniques, we can noninvasively collect neural responses during the cognitive task. One of the technical challenges is how to decode task-relevant brain states from neural signals, allowing us to explain the brain's computation and function~\cite{ito2020discovering}, especially to localize the brain networks involved in the cognitive process for deepening our understanding of cognitive impairment~\cite{kapogiannis2011disrupted}.

There are two potential ways to address the ``brain decoding'' problem: statistical analysis and machine learning. The former is to find which brain regions are activated when performing one task compared to another task or the resting state brain networks~\cite{dehaene1998inferring}. 
It relies on the assumption that once these brain regions are activated, the brain is inferred to be in the corresponding task-related brain states. However, such an assumption is not always correct, since some brain regions activated by different tasks may overlap. For example, studies have reported that the emotion and memory tasks both activate the hippocampus~\cite{drobyshevsky2006rapid,miller2009unique}. Another drawback is that not all tasks can activate certain brain regions significantly, where the activated areas we obtain greatly depend on the task design and the number of subjects.
The latter is to directly use machine learning (ML) to classify different tasks based on neural signals (\eg fMRI), regardless of whether the brain region is activated or not~\cite{woo2017building}. The performance of ML-based brain decoding relies on the selection of the model and the amount of training data. In addition, explainable ML models with high generalizability are invaluable.

Deep learning, such as convolution neural network (CNN), is an effective and scalable tool to distinguish the patterns in many tasks without manually feature selection. Recently, CNN has been brought to the brain decoding task, which uses a 3D CNN-based model to classify $7$ tasks from fMRI data~\cite{wang2020decoding}. Although CNN works well for the grid-like input in Euclidean space, such as natural images, it might not be suitable for non-Euclidean space data, especially for brain images. As we know, the neural activity between certain distant brain regions can be highly correlated, especially in the resting-state brain networks~\cite{bassett2017network}. Therefore, the geometry distance in Euclidean space may not well capture the functional distance between brain regions~\cite{rosenbaum2017spatial}. 
To better tackle non-Euclidean data types (\textit{e.g.}, graphs), some geometric deep learning methods have been proposed, such as graph convolutional networks (GCNs)~\cite{kipf2016semi} and graph attention networks (GATs)~\cite{velivckovic2018graph}.
GCN has been applied to brain decoding, which models the brain as a graph, treating the regions of interest (ROIs) as nodes and their functional connectivity as edges~\cite{li2021braingnn,zhang2021functional}. However, previous GCN applications in brain decoding studies have only focused on processing the spatial correlation between different ROIs. In order to incorporate temporal dependencies into GCN, spatio-temporal graph convolution networks (ST-GCN) have been developed and applied to traffic forecasting~\cite{yu2018spatio}. So far, its application to brain decoding has not been explored.

The motivation of our network design for brain decoding is as follows. Substantial evidence has verified that the brain is an anatomically~\cite{bullmore2009complex,bullmore2012economy} and functionally~\cite{buckner2019brain,deco2021revisiting} hierarchical organization. For example, the neural activities from different timescales simultaneously track abstract linguistic structures at distinct hierarchical levels~\cite{ding2016cortical}. The previous feature pyramid network~\cite{lin2017feature} follows such hierarchical architecture from a spatial aspect and plays a key role in the imaging processing field. Based on the above evidence and attempt, we suggest a strong and novel prediction about the relationship between network structure and brain organization for graph neural networks: the graph neural network can be improved by formulating its spatio-temporal processing module into a hierarchical structure. Hereafter, we will refer this as the ``\textit{brain-like network architecture hypothesis}''.

In this study, we propose a brain-inspired graph network architecture, STpGCN, to decode cognitive tasks. In particular, we define a spatial-temporal pathway to extract spatial and multi-scale temporal information, as well as a bottom-up pathway for the fusion of spatio-temporal dependencies at various scales. Attribute to the pyramid structure, STpGCN can better combine multi-scale temporal features of spatially distributed brain regions. 
Furthermore, by proposing an innovative model-agnostic explanation tool (\textit{BrainNetX}), we uncover the task-related brain regions from the brain-network perspective. To verify our proposed methods, we conduct extensive experiments on HCP S1200 benchmark~\cite{van2013wu}. 
STpGCN is compared with machine learning methods and a pool of the existing state-of-the-art graph models.
Our results demonstrate that STpGCN reliably outperforms the state-of-the-art models among all task designs regardless of the selection of brain atlas and fMRI time lengths. For the task-related brain regions annotation, BrainNetX with the STpGCN model can achieve better brain decoding compared with Neurosynth, an online meta-fMRI analysis tool~\cite{yarkoni2011large}. Overall, quantitative comparisons support the soundness of the ``brain-like network architecture hypothesis'' within STpGCN and the consideration of brain-network level explanation in BrainNetX.

\begin{figure*}[htbp]
\centering
\includegraphics[width=7in]{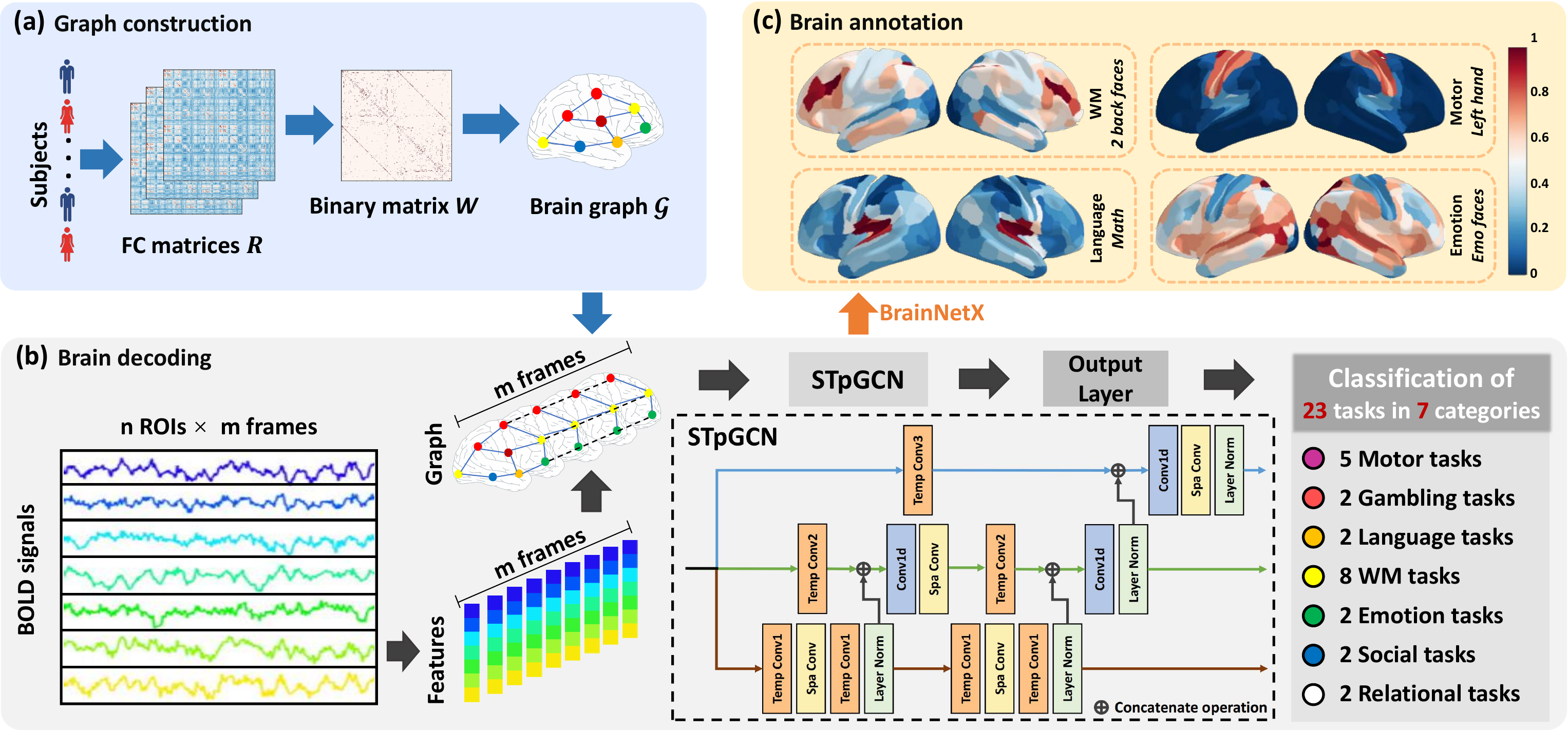}
\caption{An overview of our framework.
\textbf{(a)}\textit{ Graph construction}: The brain is parcellated into $N$ ROIs by brain atlas, and the functional connectivity (FC) between ROIs are calculated by Pearson correlation on resting-state fMRI for each individual. The brain graph $\mathcal{G}$ is constructed with k-nearest neighbors (k-NN) which performs on the binarized averaged functional connectivity $W$.
\textbf{(b)}\textit{ Brain decoding}: STpGCN takes the preprocessed parcellated individual BOLD signals time series from fMRI as input to classify 23 task-related brain states. 
\textbf{(c)}\textit{ Brain annotation:} BrainNetX together with the well-trained STpGCN annotate the core brain regions involved in each task-related brain state.} \label{fig:frame}
\end{figure*}

\section{Method}\label{sec:method}

\subsection{Problem formulation}~\label{subsec:problem_formulation}
Decoding brain states is a typical classification problem. Specifically, we study the fMRI-based task decoding problem in this study, which is to classify cognitive tasks according to the fMRI time series. We denote the labeled fMRI data $\mathcal{D} = \{(\boldsymbol{x}_i, y_i)\}_{i=1}^{n}$ as our data set, where $\boldsymbol{x}_i \in \mathbb{R}^{N\times T}$ denotes the task-related fMRI time series with $N$ regions of interests (ROIs) and $T$ time samples, $y_i$ is the task label from label set $Y$, and $n$ is the number of data. Our goal is to learn an optimal graph network $P(\cdot)$ from the training dataset and predict the task label $\hat{y}_i$ given each fMRI data $\boldsymbol{x}_{i}$ from the testing dataset,
\begin{equation} \label{eq:problem_difinition}
    \hat{y}_i= \arg\max_{y_i\in Y} P(y_i|\boldsymbol{x}_{i}).
\end{equation}

\subsection{Overall structure}~\label{subsec:overall}
The overall structure of our proposed explainable brain decoding approach is illustrated in Fig.~\ref{fig:frame}. 
Specifically, the framework consists of three modules: 

\begin{itemize}
    \item \textit{Brain graph construction} module for establishing a resting-state fMRI-based brain graph.
    \item \textit{Brain decoding} module (STpGCN) for decoding brain states via task-related fMRI data.
    \item \textit{Brain annotation} module (BrainNetX) for annotating core brain regions in each task, which explains the decoding results from a brain-network perspective.
\end{itemize}

\subsection{Brain graph construction}~\label{subsec:graph_construction}

To construct the brain graph $\mathcal{G}$, the brain is parcellated into $N$ ROIs by using the standard brain atlas. 
Here we utilize resting-state fMRI to form the brain graph $\mathcal{G}=(\mathcal{V},\mathcal{E})$,
since there is evidence that resting-state functional connectivity is thought to be a good indicator of the structural connectome of the human brain to some extent~\cite{van2009functionally,hermundstad2013structural}. 
The nodes $\mathcal{V}=\{v_1, \dots, v_N \}$ are defined as ROIs, and the edges $\mathcal{E}$ are the connectivity between nodes. 
Specifically, the brain activity 
in an ROI at time point $t$ is averaged across all voxels within the ROI.
Given $M$ subjects in total, the functional connectivity matrix of the subject $i$, $\boldsymbol{R}_i\in \mathbb{R}^{N\times N}$, is obtained by Pearson correlation of resting-state fMRI between ROIs. 
We average the adjacency matrices across all subjects to obtain the group-level adjacency matrix,
\begin{align}
    \boldsymbol{A}=\frac{1}{M}\sum_{i\in M}\boldsymbol{R}_i\in \mathbb{R}^{N\times N}.
\end{align}
Finally, the k-nearest neighbors (k-NN) strategy is applied to build up connections for each ROI with top $k$ correlated ROIs using undirected graph edges. In other words, we prune the connection from the group-level adjacency matrix $\boldsymbol{A}$, resulting in a binarized undirected group-level adjacency matrix $\boldsymbol{W}$ as the brain graph

\begin{equation}
  \boldsymbol{W}_{i,j} = \left\{ 
\begin{array}{l l}
  1, & \mbox{if} ~\boldsymbol{A}_{i,j} \geq k\text{th-largest }(\boldsymbol{A}_{i}), \forall \boldsymbol{A}_{i,j}\in \boldsymbol{A}_i\\
  0, & \mbox{else}.\\ \end{array} \right.
\label{incidenceMatrix}
\end{equation}


\subsection{STpGCN for brain decoding}
\label{subsec:stpgcn}
The key idea of the proposed method is to leverage multi-scale features behind raw data, which has semantics from low to high levels and construct a spatial temporal-pyramid feature with multi-level semantics throughout. To this end, we introduce a novel \textit{Spatial Temporal-pyramid Graph Convolutional Network} (STpGCN) that shares a spatial graph convolution layer and temporal convolution layer with the state-of-the-art temporal information forecasting graph networks~\cite{yu2018spatio}. The STpGCN takes multi-channel time series (of size $>$ 4) as input and outputs a spatial temporal-pyramid feature. Our architecture has three spatial-temporal pathways with different-scale temporal convolutions and several bottom-up pathways, as described in the following.

\subsubsection{Spatial-temporal pathway}
The spatial-temporal pathway aims at pursuing distinct representations along the various temporal dimensions while also taking spatial feature correlation into account. The STpGCN contains three spatial-temporal pathways. Each of them is composed of temporal convolution modules and spatial convolution modules.

\textbf{Spatial Graph Convolution Module:}
The notion of spatial graph convolution (Spa Conv) operator $*_\mathcal{G}$ can be defined as the multiplication of an input vector $\boldsymbol{x}$ with a filter $\Theta(\cdot)$,
\begin{equation}
    \Theta *_{\mathcal{G}} \boldsymbol{x}=\Theta(\boldsymbol{L}) \boldsymbol{x}=\Theta\left(\boldsymbol{U} \boldsymbol{\Lambda} \boldsymbol{U}^{T}\right) \boldsymbol{x}=\boldsymbol{U} \Theta(\boldsymbol{\Lambda}) \boldsymbol{U}^{T} \boldsymbol{x},~\label{eq:spectral}
\end{equation}
where $\boldsymbol{U}$ and $\boldsymbol{\Lambda}$ are the eigenvector matrix and the diagonal matrix of eigenvalues of the normalized graph Laplacian $\boldsymbol{L} \in \mathbb{R}^{N \times N}$, respectively. We have $\boldsymbol{L}=\boldsymbol{U} \boldsymbol{\Lambda} \boldsymbol{U}^{T}$.

The graph Laplacian $\boldsymbol{L}$ is from the transformation of the adjacency matrix $\boldsymbol{W}$ with $\boldsymbol{L} = \boldsymbol{I}_N - \boldsymbol{D}^{-\frac{1}{2}} \boldsymbol{W} \boldsymbol{D}^{-\frac{1}{2}}$, where $\boldsymbol{D}$ is the diagonal degree matrix with $\boldsymbol{D}_{i,i}=\sum_{j}{\boldsymbol{W}}_{i,j}$, and $\boldsymbol{I}_N$ is an identity matrix.
In this study, the adjacency matrix $\boldsymbol{W}\in \mathbb{R}^{N\times N}$ is obtained from fMRI data. 

To reduce the computational complexity, Chebyshev polynomial $T_k(\cdot)$ is commonly applied to approximate the filter $\Theta$ ~\cite{hammond2011wavelets,defferrard2016convolutional}. Thus, the spatial graph convolution can be rewritten as a linear function of Chebyshev polynomial $T_k(\tilde{\boldsymbol{L}})$,
\begin{equation}
\Theta *_{\mathcal{G}} \boldsymbol{x} = \Theta(\boldsymbol{L})\boldsymbol{x}\approx\sum_{k=0}^{K}\theta_{k}T_{k}(\tilde{\boldsymbol{L}})\boldsymbol{x},~\label{eq:chebyshev}
\end{equation}
where $\tilde{\boldsymbol{L}} = 2\boldsymbol{L}/\lambda_{max}-\boldsymbol{I}_{N}$, and $\lambda_{max}$ denotes the largest eigenvalue of $\boldsymbol{L}$.

By introducing the first-order approximation~\cite{kipf2016semi}, we set that $\lambda_{max}\approx2$ and $\theta = \theta_0 = -\theta_1$. Substituting $\tilde{\boldsymbol{W}}=\boldsymbol{W}+\boldsymbol{I}_N$ and $\tilde{\boldsymbol{D}}_{ii}=\sum_j\tilde{\boldsymbol{W}}_{ij}$ into Eq.~\eqref{eq:chebyshev}, we can approximate the spatial graph convolution as
\begin{equation} \label{eq:Spa-Conv-simplied}
    \begin{aligned}
     \Theta *_{\mathcal{G}} \boldsymbol{x} &\approx \theta_{0} \boldsymbol{x}-\theta_{1}\left(\boldsymbol{D}^{-\frac{1}{2}} \boldsymbol{W} \boldsymbol{D}^{-\frac{1}{2}}\right) \boldsymbol{x}\nonumber\\
    &\approx\theta\left(\boldsymbol{I}_{N}+\boldsymbol{D}^{-\frac{1}{2}} \boldsymbol{W} \boldsymbol{D}^{-\frac{1}{2}}\right) \boldsymbol{x} \nonumber\\
    &\approx\theta\left(\tilde{\boldsymbol{D}}^{-\frac{1}{2}} \tilde{\boldsymbol{W}} \tilde{\boldsymbol{D}}^{-\frac{1}{2}}\right) \boldsymbol{x}.\nonumber
    \end{aligned}
\end{equation}

\textbf{Temporal Convolution Module}:
The design of temporal convolution (Temp Conv) contains a 1-D convolution with a width $K_{j}$ kernel followed by a gated linear unit (GLU) to add non-linearity. Here, $K_{j}$ denotes temporal convolution kernel in $j$-th spatial-temporal pathway (Detailed in Section~\ref{subsec:stpgcn}). The input of each node in the graph can be regarded as $\boldsymbol{u}\in \mathbb{R}^{l\times c_i}$, where $l$ denotes input length, $c_i$ denotes input channel. The temporal convolution kernel $\mathcal{K}\in\mathbb{R}^{K_{j}\times c_i \times 2c_o}$ performs to map $\boldsymbol{u}$ to a single output element $[\boldsymbol{Z} \boldsymbol{V}]\in\mathbb{R}^{(l-K_{j}+1)\times(2c_o)}$ ($[\boldsymbol{Z} \boldsymbol{V}]$ is split in half with the same size of channels, $c_o$ denotes output channel). Accordingly, the temporal gated convolution module can be denoted as,
\begin{equation}
    \mathcal{K} *_{\mathcal{T}} \boldsymbol{u} = \boldsymbol{Z} \odot \sigma(\boldsymbol{V}) \in \mathbb{R}^{(l-K_{j}+1)\times c_o},
\end{equation}
where $*_{\mathcal{T}}$ denotes temporal convolution; $\boldsymbol{Z}$ and $\boldsymbol{V}$ are the input of gates in GLU, respectively; $\odot$ denotes the element-wise Hadamard product; $\sigma(\boldsymbol{V})$ denotes the sigmoid gate which determines the importance of input $\boldsymbol{Z}$.

The spatial convolution module in the spatial-temporal pathway can be any spatial graph convolutional model (\textit{e.g.,}~\cite{kipf2016semi,velivckovic2018graph}) that performs on time series data with graph structure. For brevity, we apply the GCN~\cite{kipf2016semi} as the spatial convolution module. Similar to the spatial convolution module, generally, there are also many choices to integrate temporal information into the temporal convolution module (\textit{e.g.,}~\cite{zhang2019view,zhang2020neural}). Despite the simple structure of CNNs, they converge quickly and perform well in many scenarios. Thus, we take 1-D convolution followed by a GLU as temporal convolution.

\subsubsection{Bottom-up pathway}
The bottom-up pathway is the concatenation of fine-grained and coarse-grained features from different spatial-temporal pathways. Specifically, the bottom spatial pathway gradually flows multi-scale information up to the top pathway to produce features with both global and local consideration, yielding multi-level semantic information. Inspired by the bottom-up processing in the brain, we term this network structure in 
our model as \textit{Bottom-up pathway}. The merit of the bottom-up pathway is to make the higher-level features contain higher-level semantics but also hold promise to retain localized information. The following 1-D convolution after each concatenation operates the concatenated features for each node, so the fused features always have $N$ rows along the node dimension, maintaining spatial fidelity as much as possible.

The state-of-the-art model, ST-GCN, is indicated as the bottom spatial-temporal pathway (brown line in Fig.~\ref{fig:frame}(b)) of STpGCN with a fixed width $K_{1}$ 1-D convolution kernel. We extend the original ST-GCN by adding two additional spatial-temporal pathways, with a width $K_{2}$ kernel (mid) and a width $K_{3}$ kernel (top) indicated by the green and blue lines in Fig.~\ref{fig:frame}, respectively. We define the size of these three temporal convolution kernels as $K_{1} = T/4$, $K_{2} = 2K_{1}+1$ and $K_{3} = \min\{4K_{1}+1, T\}$. A ReLU function is used as the activation after each spatial convolution to enhance non-linearity.

\subsection{BrainNetX for annotating tasks-related brain regions}~\label{subsec:BrainNetX}
The details of BrainNetX are as follows. Formally, let $P(\cdot)$ denote a well-trained graph classification model. Given task-related brain graph $\mathcal{G}$ with $\boldsymbol{x}$ as input, the objective of our explanation task is to measure the importance of each ROI for each brain decoding task from the brain-network aspect. For each task-relevant fMRI data $\boldsymbol{x}_i$, the algorithm first assigns $N$ ROIs into $Q$ distinct brain networks and transforms the original ROI-wise dataset into brain-network-wise dataset, $F_{\boldsymbol{x}_i}=\{f_{\boldsymbol{x}_{i, 1}}, \dots,  f_{\boldsymbol{x}_{i, Q}}\}$, where each brain network $q$ of $\boldsymbol{x}_i$, $f_{\boldsymbol{x}_{i, q}}\in R^{N_q\times T}$, has $T$ time points and $N_q$ ROIs ($N_q<N$, $\forall q\in\{1, \dots, Q\}$).
Then, a masking strategy and a keeping strategy are applied to the brain networks. Specifically, the masking strategy is to set all the node features inside the mask to zeros, while the keeping strategy is to set the node features outside the mask to zeros. Accordingly, the contribution of each brain network for brain decoding can be defined as follows. For each brain network $q$ of $\boldsymbol{x}_i$, let $\phi_{mask}(f_{\boldsymbol{x}_{i, q}}) = P(F_{\boldsymbol{x}_i}) - P(F_{\boldsymbol{x}_i}\backslash \{f_{\boldsymbol{x}_{i, q}}\})$ and $\phi_{keep}(f_{\boldsymbol{x}_{i, q}}) = P(f_{\boldsymbol{x}_{i, q}})$ be the importance of $f_{\boldsymbol{x}_{i, q}}$ derived from masking strategy and keeping strategy, respectively. Finally, we take the perturbed features by the above two strategies as input into the well-trained model and obtain the \textit{importance score} $\mathcal{I}({f_{\boldsymbol{x}_{i, q}}})$ of each brain network $f_{\boldsymbol{x}_{i, q}}$, which be represented as
\begin{align}
    \mathcal{I}({f_{\boldsymbol{x}_{i, q}}}) = \sigma(\alpha(\bar{\phi}_{mask}(f_{\boldsymbol{x}_{i, q}}) + (1-\alpha) \bar{\phi}_{keep}(f_{\boldsymbol{x}_{i, q}})))
\end{align}
where $\bar{\phi}_{mask}$ and $\bar{\phi}_{keep}$ denote $\phi_{mask}$ and $\phi_{keep}$ after normalization, respectively; $\alpha$ and $1-\alpha$ separately denote the weight proportions of masking strategy and keeping strategy; $\sigma$ denotes normalization.

It is worth noting that in some situations the graph might be sensitive to structural perturbation~\cite{schlichtkrull2021interpreting}. In our case, BrainNetX only perturbs the node features, rather than distorting the graph structure, as the brain graph is defined based on the well-known brain atlas. Furthermore, BrainNetX is a model-agnostic method that is able to provide whole-brain explanation given any task-related brain states fMRI decoded by any machine learning model or deep learning model.

\section{Experiments and results}
\label{sec:experiments_analysis}

\subsection{Dataset}

In this work, we utilize minimally preprocessed resting-state fMRI and task-evoked fMRI data from the HCP S1200 Release~\footnote{https://humanconnectome.org/study/hcp-young-adult} dataset~\cite{van2013wu} to validate our model. 
The HCP S1200 fMRI dataset contains 1200 subjects, most of whom performed all $23$ tasks across $7$ categories, including working memory (WM), gambling, motor, language, social cognition, relational processing and emotion processing tasks.

\subsection{Alternative baselines}
So far the brain decoding problem has rarely been studied using GNN models. Here, we implement four GNN models (\ie GCN, GIN, GAT, ST-GCN) as baselines to verify the capability of STpGCN. We also add two widely used models, SVM-RBF and MLP-Mixer, as the representatives of traditional machine learning models and deep learning models, respectively.

\textbf{SVM-RBF}~\cite{melgani2004classification} indicates SVM with radial basis function (RBF) kernel, which is a representative classification method.

\textbf{MLP-Mixer}~\cite{tolstikhin2021mlp} is a state-of-the-art deep learning model for computer vision and natural language processing. We employ it here as a competing model in Euclidean space. 

\textbf{GCN}~\cite{kipf2016semi} introduces simplifications into graph convolutions known from spectral graph theory to define parameterized filters similar to classical convolutional neural networks. 

\textbf{GIN}~\cite{xu2018powerful} generalizes the Weisfeiler-Lehman (WL) graph isomorphism test and performs graph convolution with maximum discriminative power among GNNs. 

\textbf{GAT}~\cite{velivckovic2018graph} leverages the attention mechanism to implicitly assign different weights to different nodes within a neighborhood. 

\textbf{ST-GCN}~\cite{yu2018spatio} is a state-of-the-art time series forecasting method, which is mainly used for traffic flow prediction. In this work, instead of predicting traffic flow, we adapt ST-GCN for decoding distinct brain states. 


\subsection{Evaluation metrics}
We use 4 evaluation metrics to validate the decoding performance of the models in our experiments: accuracy (ACC), macro precision score (Macro pre), macro recall (Macro R), and macro F1 score (Macro F1). The higher these metrics, the better the performance.
\begin{table}[htbp]
    \centering
    \caption{Details of HCP task-fMRI dataset after preprocessing. }\label{tab:datasets}
    \setlength{\tabcolsep}{1pt}
    \begin{tabular}{cccc}
        \hline
        \textbf{Category} & \textbf{\makecell[c]{$\#$ of\\ subjects}} & \textbf{\makecell[c]{$\#$ of time points}} & \textbf{$\#$ tasks} \\ \hline
        Working memory & 1085   & \{4, 7, 10, 13, 15\}& 8\\
        Gambling & 1080  & \{4, 7, 10, 13, 15\} & 2\\
        Motor & 1083   & \{4, 7, 10, 13, 15\} & 5\\
        Language & 1051  & \{4, 7, 10, 13, 15\} & 2\\
        Social cognition & 1051  & \{4, 7, 10, 13, 15\} & 2\\
        Relational processing & 1043 & \{4, 7, 10, 13, 15\} & 2\\
        Emotion processing & 1047  & \{4, 7, 10, 13, 15\} & 2\\
        \hline
    \end{tabular}
\end{table}

\subsection{Experimental setup}
We segmented the task-evoked fMRI data from the HCP dataset according to event triggers in each trial ($T=\{4, 7, 10, 13, 15\}$ time points) for each subject.
To avoid the class imbalance problem, we balanced the data in each task by sampling a single trial from each subject. The preprocessed dataset is detailed in Table~\ref{tab:datasets}. 

To test whether the results depend on the way to parcellate the brain, two brain atlases are employed to define the ROIs for graph construction: (1) automated anatomical labeling (AAL) atlas~\cite{tzourio2002automated} with the number of regions $N = 90$, (2) multi-modal parcellation (MMP) atlas~\cite{glasser2016multi} with the number of regions $N = 360$. These ROIs can be grouped into $17$ cortical networks following yeo17 atlas~\cite{thomas2011organization}.
To tune the hyperparameter, we examine the decoding performance under $k\in\{1,\dots,5\}$ for kNN in the validation set. Finally, we set $k$=5 for the test dataset, to achieve high decoding accuracy with sufficient sparsity of the brain graph.

In STpGCN, the learnable parameters are $\theta$ and $K_t$.
In BrainNetX, the hyperparameter $\alpha$ is set to $0.5$. After hyperparameter tuning, for a fair comparison, all deep graph models are composed of 2 layers and trained under the same conditions, batch size = 64, epochs = 64, Adam optimizer with an initial learning rate of $0.001$ and decay the learning rate by $0.5$ after the validation loss plateaued (patience = $5$). For MLP-Mixer, we set patch size = 5, dimensions of hidden layer = 512, dropout = 0.5. For SVM, we set the maximum iteration number = 64, regularization parameter = 1. 
For each configuration, we perform a $10$-fold cross-validation and report the mean and standard deviation (std) of validation accuracy. 
We use the cross-entropy loss as the optimization metric. 
Experiments are implemented by PyTorch $1.9.0$ and DGX $0.7.1$, and trained with NVIDIA GeForce RTX $3080$ GPU.

\begin{table*}[ht]
    \centering
    \begin{threeparttable}[b]
	\caption{Comparison of the $23$ task-related brain states decoding results ($\%$) of our model and baseline models on $15$ time points of fMRI with MMP atlas and AAL atlas across $10$-fold cross-validation. }\label{tab:deco_result}
    \setlength{\tabcolsep}{1pt}
	\small
	\begin{tabular}{c|cccc|cccc|ccc}
	\toprule
		\multirow{1}{*}{\diagbox{Methods}{Metrics}} & 
		\multicolumn{4}{c|}{AAL Atlas} &
		\multicolumn{4}{c|}{MMP Atlas} &
		\multicolumn{3}{c}{Modules}\\
		& ACC$\uparrow$    & Macro Pre$\uparrow$ & Macro R$\uparrow$ & Macro F1$\uparrow$ & ACC$\uparrow$    & Macro Pre$\uparrow$ & Macro R$\uparrow$ & Macro F1$\uparrow$ & Mid STP & Top STP & BUP\\
		\midrule
        (a) \underline{\textbf{Baselines:}} &&&&&&&&&&&\\
        SVM-RBF & 57.9$\pm$1.8  & 63.7$\pm$1.7     &   57.9$\pm$1.8  & 57.8$\pm$1.9 & 79.9$\pm$1.6  & 81.8$\pm$1.3     &   80.0$\pm$1.6  & 80.0$\pm$1.6  &&&  \\
        MLP-Mixer & 81.7$\pm$1.4  & 82.4$\pm$1.2     &   81.7$\pm$1.4  & 81.6$\pm$1.4 & 89.8$\pm$1.0  & 89.9$\pm$0.9     &   89.8$\pm$0.9  & 89.8$\pm$0.9  &&& \\
        GCN & 37.4$\pm$1.4  & 36.2$\pm$1.3     &   37.4$\pm$1.3  & 36.3$\pm$1.3 & 61.4$\pm$1.6  & 60.8$\pm$1.5     &   61.4$\pm$1.6  & 60.7$\pm$1.5  &&& \\
        GIN  & 64.1$\pm$2.7  & 64.6$\pm$3.0     &   64.2$\pm$2.7  & 63.7$\pm$2.8 & 80.1$\pm$2.1  & 80.6$\pm$2.1     &   80.1$\pm$2.1  & 80.1$\pm$2.1 &&&  \\
        GAT & 64.1$\pm$2.7  & 64.6$\pm$3.0     &   64.2$\pm$2.7  & 63.7$\pm$2.8 & 82.2$\pm$1.2  & 82.3$\pm$1.2  &  82.3$\pm$1.3   &   82.2$\pm$1.3 &&&\\
        ST-GCN  & 74.9$\pm$1.8  & 75.1$\pm$1.7     &  75.0$\pm$1.8   &   74.8$\pm$1.8 & 90.6$\pm$0.6  & 90.7$\pm$0.6     &  90.6$\pm$0.6   &   90.6$\pm$0.6 &&&\\
        \midrule
        \underline{\textbf{Proposed:}} &&&&&&&&&&&\\
        (b) STpGCN-$\alpha$ & 81.8$\pm$2.2  & 82.0$\pm$2.1     &  81.8$\pm$2.1   &   81.8$\pm$2.2 & 91.0$\pm$0.6  & 91.1$\pm$0.6     &  91.0$\pm$0.6   &   91.0$\pm$0.6 &&$\checkmark$&$\checkmark$\\
        (c) STpGCN-$\beta$  & 81.4$\pm$1.5  & 81.5$\pm$1.5     &  81.3$\pm$1.5   &   81.3$\pm$1.5 & 91.2$\pm$0.8  & 91.3$\pm$0.8     &  91.3$\pm$0.8   &   91.0$\pm$0.8 &$\checkmark$&&$\checkmark$\\ 
        (d) STpGCN-$\gamma$ & 80.3$\pm$2.6  & 80.4$\pm$2.6     &  80.3$\pm$2.6   &   80.2$\pm$2.6 & 90.7$\pm$0.7  & 90.9$\pm$0.7     &  90.7$\pm$0.7   &   90.7$\pm$0.7 &$\checkmark$&$\checkmark$&\\ 
        (e) STpGCN & \textbf{82.3$\pm$1.6}  & \textbf{82.5$\pm$1.6}     & \textbf{82.3$\pm$1.6}    & \textbf{82.3$\pm$1.6} & \textbf{92.4$\pm$0.8} & \textbf{92.5$\pm$0.7}     & \textbf{92.4$\pm$0.8}    & \textbf{92.4$\pm$0.7}  &$\checkmark$&$\checkmark$& $\checkmark$\\
        \bottomrule
	\end{tabular}
	\end{threeparttable}
    \begin{tablenotes}[flushleft]
     \item Note: STP and BUP denote spatial-temporal pathway and bottom-up pathway, respectively. Mean $\pm$ std are from 10-fold cross-validation. The best results are highlighted in bold.
   \end{tablenotes}
\end{table*}

\begin{figure*}[ht]
\centering
\includegraphics[width=.9\textwidth]{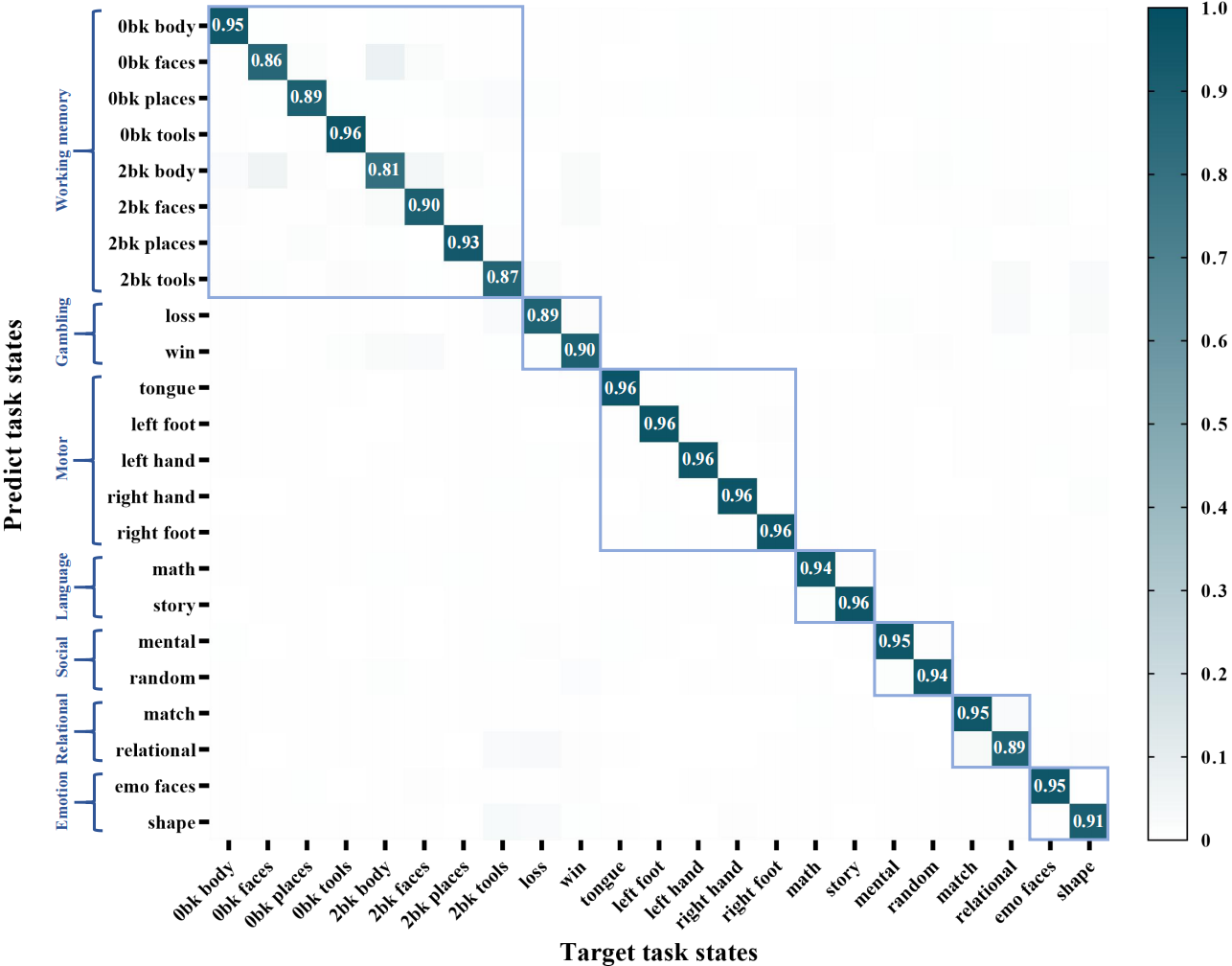}
\caption{Confusion matrix of $23$ task-related brain states classification on $15$ time points of fMRI with MMP atlas using STpGCN.} \label{fig:confusion_mat}
\end{figure*}

\subsection{Brain decoding performance} \label{Res:decoding}

\subsubsection{Comparison to state-of-the-art}

We first validate the proposed biologically inspired brain decoding method, STpGCN, for brain decoding by comparing classification performance with other models using task-related fMRI. Models with a higher capability of brain decoding should have higher accuracy in the testing set. Table~\ref{tab:deco_result}(a, e)  summarizes the comparison results between STpGCN with other baselines. The results demonstrate that STpGCN outperforms all baselines and the state-of-the-art models in terms of four metrics (Macro F1 score = 82.3$\pm$1.6 and 92.4$\pm$0.7 using AAL atlas and MMP atlas, respectively). More importantly, STpGCN achieves better performance compared to ST-GCN by about 7.5$\%$ and 1.8 $\%$ based on the Macro F1 score, indicating that the pyramid design is indeed helpful for brain decoding. 

To examine the classification accuracy of each task, the confusion matrix of 23 tasks from STpGCN using the MMP atlas is illustrated in Fig.~\ref{fig:confusion_mat}. It is shown that STpGCN produces an excellent block diagonal confusion matrix, indicating all $23$ tasks can be identified with high accuracy (ranging from 81\% to 96\%). In all categories, the average decoding accuracy is promising (average identification rate = 0.896 / 0.895 / 0.960 / 0.950 / 0.945 / 0.920 / 0.930 for working memory, gambling, motor, language, social relational and emotion, respectively).
The confusion matrix of 23 tasks from STpGCN using AAL atlas also demonstrates satisfactory results (average identification rate = 0.775 / 0.760 / 0.880 / 0.925 / 0.885 / 0.660 / 0.800 for working memory, gambling, motor, language, social relational and emotion, respectively; Supplementary Fig.~S1).

%


\subsubsection{Ablation study}

In the ablation study, we remove the spatial-temporal pathway layer by layer. Table~\ref{tab:deco_result}(b, c) shows the STpGCN without top spatial-temporal pathway (\ie STpGCN-$\alpha$) and middle spatial-temporal pathway (\ie STpGCN-$\beta$), respectively. With the latter modification, the output features by the last layer normalization module in the bottom spatial-temporal pathway are concatenated with the output features by the temporal convolution module in the top spatial-temporal pathway. These two structures simulate the effect of using different pyramidal spatial-temporal graph architectures.

The results reveal that the STpGCN-$\alpha$  and STpGCN-$\beta$ are better than the ST-GCN, but inferior to STpGCN with a full pyramidal structure (Table~\ref{tab:deco_result}(e)).
The results of STpGCN without a bottom-up pathway (\ie STpGCN-$\gamma$) are presented in Table~\ref{tab:deco_result}(d). We find that this variant outperforms the baselines while its performance is worse than the above two variants (\ie STpGCN-$\alpha$ and STpGCN-$\beta$) which only have two spatial-temporal pathways. A detailed description of the network architectures of STpGCN-$\alpha$, STpGCN-$\beta$ and STpGCN-$\gamma$ are provided in Supplementary Fig.~S2.

\begin{table}[ht]
    \centering
    \setlength{\tabcolsep}{2pt}
    \caption{Comparison results of $23$ brain states decoding using STpGCN on $15$ time points of fMRI with MMP atlas given different dataset split ratios. 
    }\label{tab:robust_diff_ratio}
    \begin{tabular}{c|cccc}
        \hline
        \makecell[c]{Training set\\ ratios}         & ACC$\uparrow$    & Macro Pre$\uparrow$ & Macro R$\uparrow$ & Macro F1$\uparrow$ \\ \hline
        90$\%$  & 92.4$\pm$0.8  & 92.5$\pm$0.7     &  92.4$\pm$0.8  &   92.4$\pm$0.7 \\
        70$\%$  & 90.7$\pm$0.6  & 90.8$\pm$0.6     &   90.7$\pm$0.6  & 90.7$\pm$0.6   \\
        50$\%$  & 88.8$\pm$0.5  & 88.8$\pm$0.5     &   88.8$\pm$0.5  & 88.8$\pm$0.5   \\
        30$\%$  & 86.4$\pm$0.3  & 86.4$\pm$0.2     &   86.4$\pm$0.3  & 86.4$\pm$0.3   \\
        10$\%$  & 79.2$\pm$0.5  & 79.3$\pm$0.4     &   79.2$\pm$0.4  & 79.2$\pm$0.5   \\
        \hline
    \end{tabular}
    \begin{tablenotes}
     \item Note: Mean $\pm$ std ($\%$) are from 10-fold cross-validation.
    \end{tablenotes}
\end{table}

\subsection{Robustness and stability of STpGCN} \label{sec:robust}

A big challenge of using deep learning for brain decoding is the small sample size of brain images.
Many existing deep learning models achieve good results only under extremely large datasets. 
An ideal brain decoding model should be robust when the training set is reduced, in other words, a decreased training set ratio only causes slight performance degradation. To examine the robustness of STpGCN with the size of training data, we vary the ratio of the training set. Table~\ref{tab:robust_diff_ratio} shows 4 evaluation metrics of decoding $23$ task-related brain states as decreasing training set ratio. The results demonstrate that STpGCN is reliable across different training sizes (Macro F1 score = 92.4$\pm$0.7  / 90.7$\pm$0.6  / 88.8$\pm$0.5  / 86.4$\pm$0.3 / 79.2$\pm$0.5 when training set ratios equals to 90$\%$, 70$\%$, 50$\%$, 30$\%$and 10$\%$, respectively).

\begin{figure}[ht]
\begin{center}
\includegraphics[width=.48\textwidth]{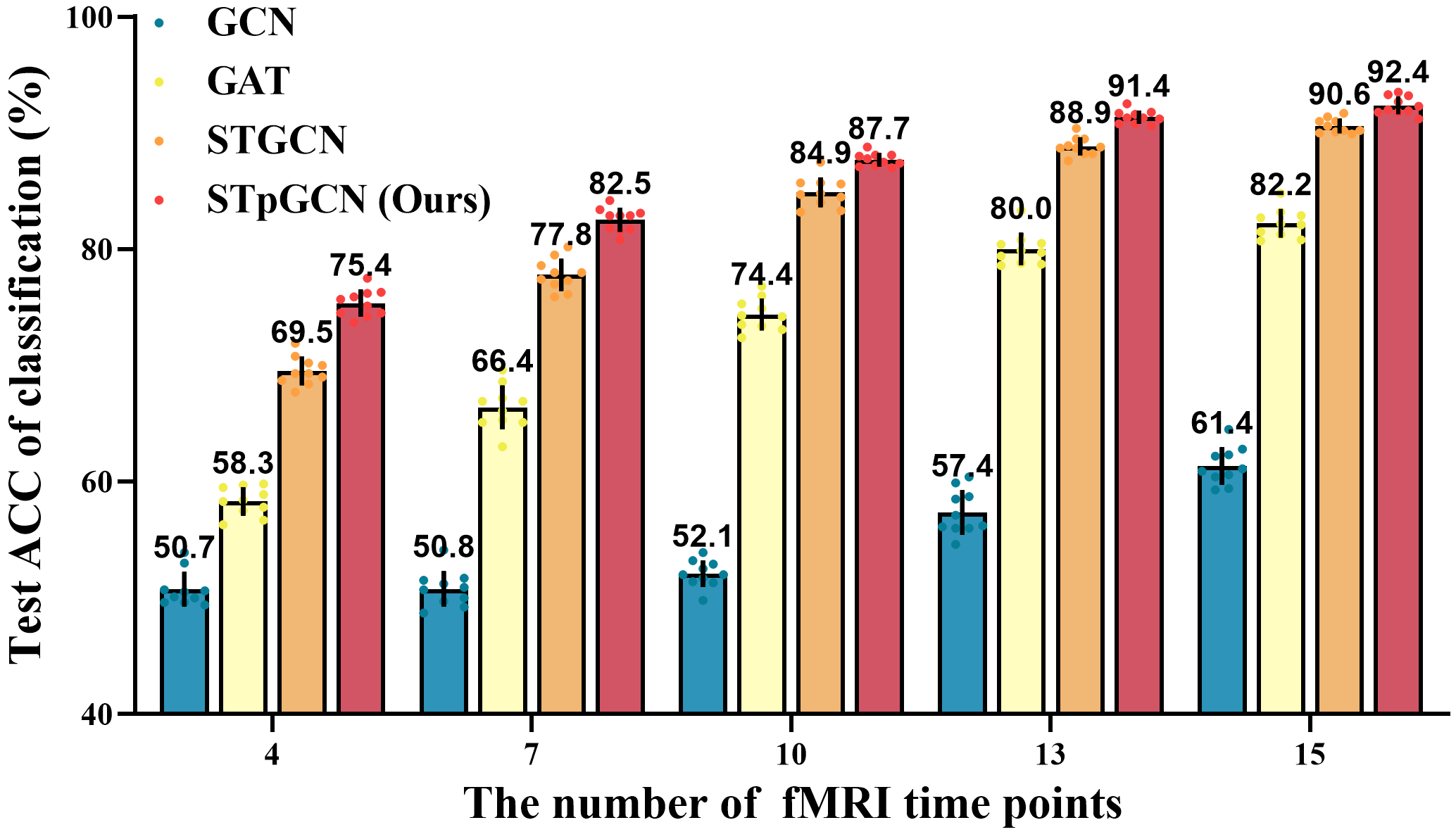}
\end{center}
\caption{Classification accuracy ($\%$) of different lengths of fMRI with MMP atlas. The time points (lengths) vary from $4$ to $15$ (repetition time Tr=$0.72$). The results of GCN, GAT, ST-GCN, and STpGCN are shown in blue, yellow, orange and red, respectively. The number and error bar indicate the mean and std of test accuracy across $10$-fold cross-validation.} \label{fig:robust_diff_time_step}
\end{figure}

We further test the model robustness under different lengths of fMRI data. We compare STpGCN with other graph models (\eg GCN, GAT and STpGCN), using 4, 7, 10, 13, and 15 fMRI time points.
The results are shown in Fig.~\ref{fig:robust_diff_time_step}.
Notably, STpGCN reliably achieves the highest accuracy in each time length. The advantage of STpGCN is more pronounced when the time window length of task-related fMRI is shorter. It is remarkable that the accuracy of decoding $23$ task-related brain states can reach $82.5\%$ with only $7$ fMRI time points.

\subsection{Explainability of STpGCN} \label{sec:explainability}

\begin{figure*}[htbp]
    \begin{center}
    \includegraphics[width=7in]{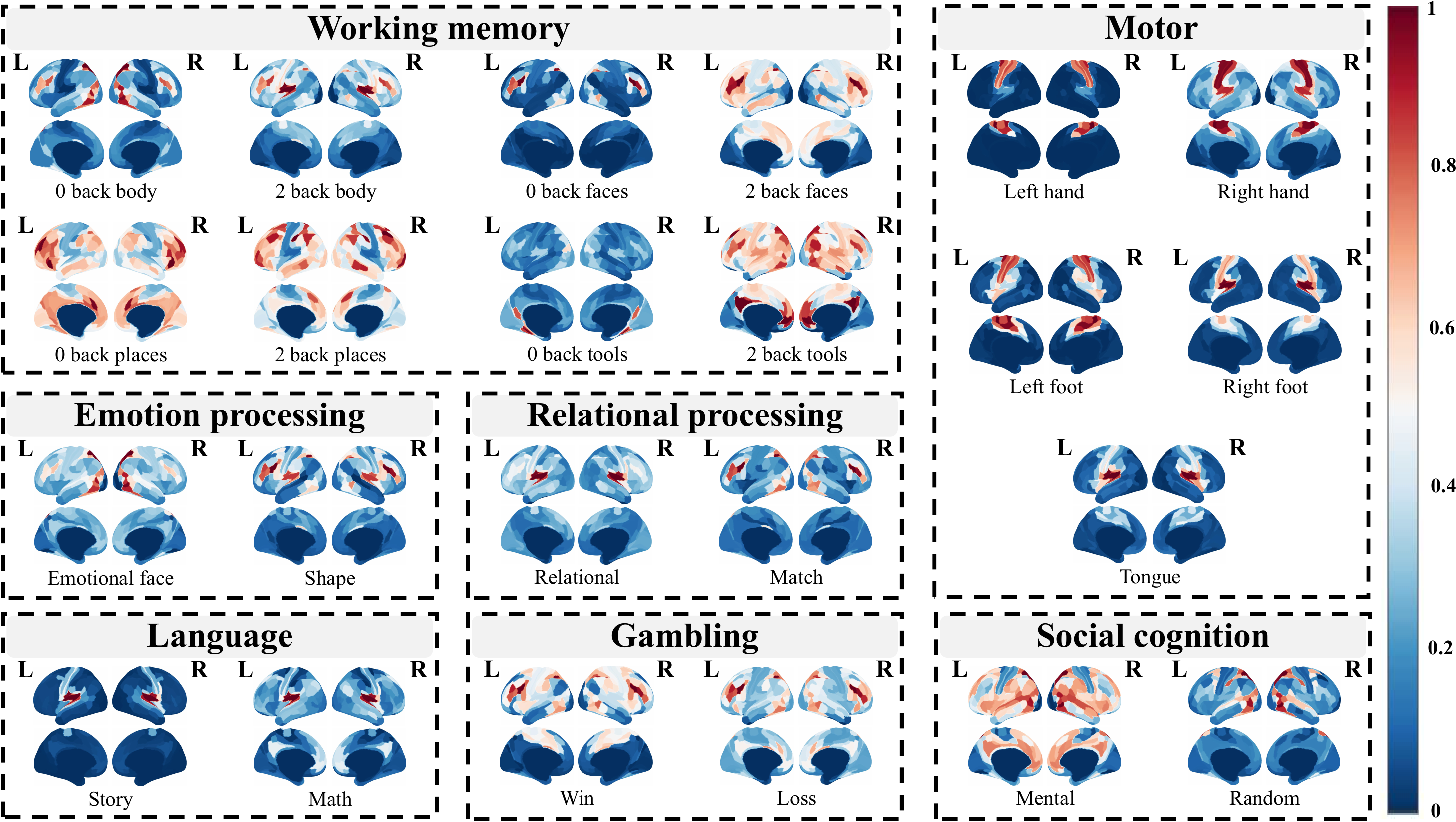}
    \end{center}

    \caption{Explanability of STpGCN. We visualize the importance score of 23 task-related brain states using STpGCN given 15 time points of fMRI with MMP atlas. The importance score is averaged across 10-fold cross-validation.}
    \label{fig:annotation_stpgcn} 
\end{figure*}

\begin{figure*}[htbp]
    \begin{center}
    \includegraphics[width=7in]{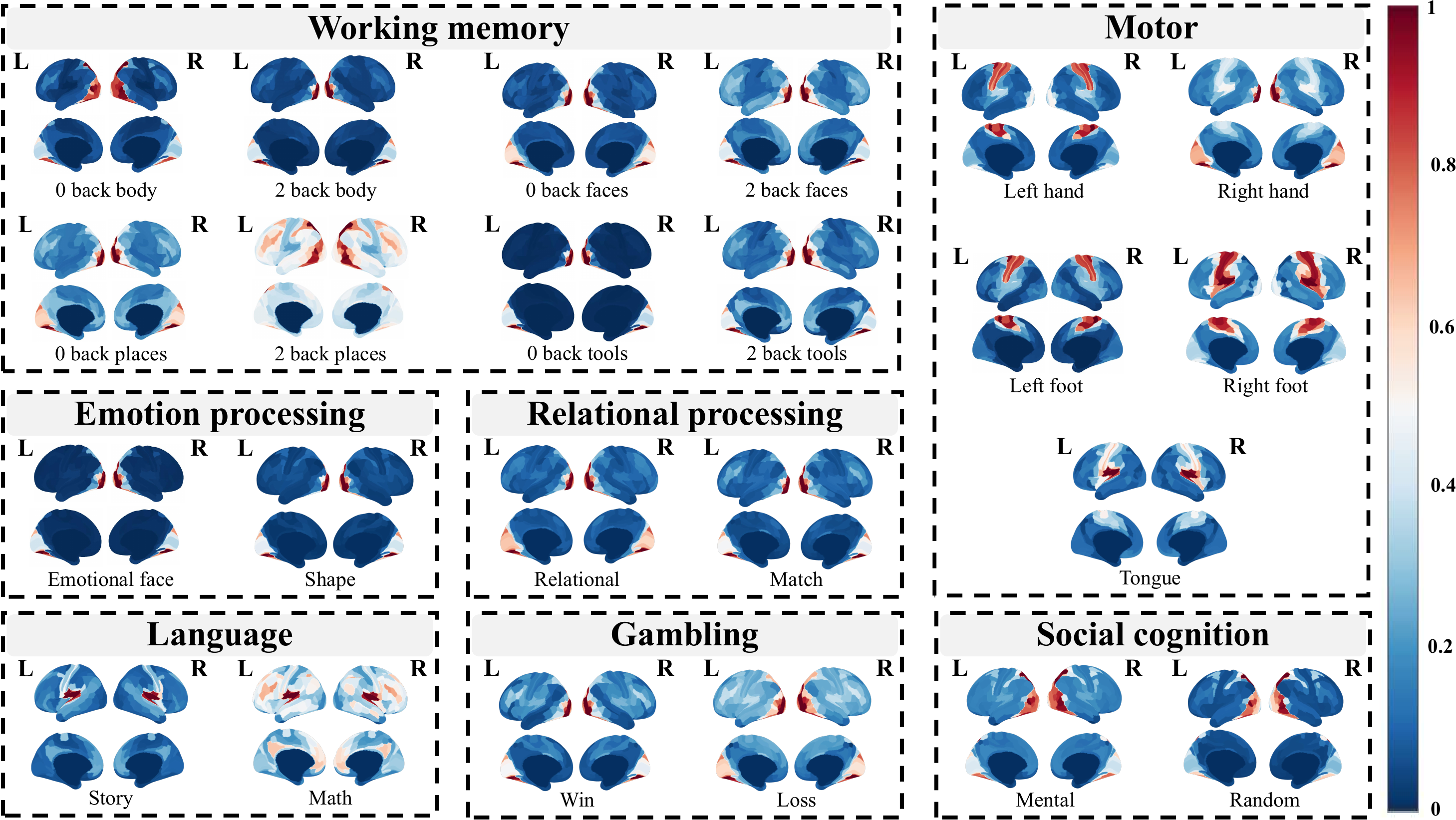}
    \end{center}
    \caption{Explanability of MLP-Mixer. We visualize the importance score of 23 task-related brain states using MLP-Mixer given 15 time points of fMRI with MMP atlas. The importance score is averaged across 10-fold cross-validation.}
    \label{fig:annotation_MLP-Mixer} 
\end{figure*}

\textbf{Visualization of task-related brain regions}: To visualize the whole-brain explanation of the STpGCN modeling results, we apply BrainNetX to annotate the core brain regions of each task. Fig.~\ref{fig:annotation_stpgcn} displays the topological maps of importance score in 23 tasks using STpGCN, and Fig.~\ref{fig:annotation_MLP-Mixer} are from MLP-Mixer. Using BrainNetX, the working memory-related brain regions annotated by STpGCN involved a wide range of regions, including the dorsolateral prefrontal cortex, the inferior parietal lobule, and the anterior lateral occipital lobe, whereas MLP-Mixer mainly annotates visual areas. Differences in brain regions annotated by STpGCN and MLP-Mixer are more pronounced on the higher-order cognitive tasks (\eg emotion process, relational processing, gambling and social cognitive tasks), and less obvious on low-order cognitive tasks (\eg motor tasks). For more comparisons, the visualization of brain annotation using ST-GCN and BrainNetX, as well as the brain regions from the state-of-the-art annotation tool, Neurosynth~\cite{yarkoni2011large}, are shown in Supplementary Fig.~S3 and Fig.~S5, respectively.



For the above annotation results of STpGCN and BrainNetX, many brain regions belong to well-known resting-state networks (RSNs)~\cite{liu2017detecting}, such as the motor network for motor tasks and the language network for language tasks, which are well in line with previous studies ~\cite{preti2019decoupling}. Notably, the core brain regions for higher-order cognitive tasks highlighted by BrainNetX and STpGCN are largely overlapped with task-related brain regions identified by cognitive neuroscientists through traditional statistical analysis; for instance, the theory of mind brain areas (\eg, temporal parietal junction and inferior and superior temporal sulcus) in social tasks~\cite{zheng2021graph}, anterior prefrontal cortex and dorsolateral prefrontal cortex for working memory tasks~\cite{curtis2003persistent}. More importantly, STpGCN can find some regions which not be significant in meta analysis while still have theoretically or experimentally evidence to support they are task relevant, such as the medial temporal lobe, lateral temporal lobe and posterior cingulate cortex for working memory tasks~\cite{glasser2016multi}, and the superior parietal lobule for social cognition tasks~\cite{glasser2016multi}.


\begin{table*}[t]
	\centering
	\begin{threeparttable}[b]
	\caption{\label{tab:top_k_roi_svm_comparison}
    Comparison of the explainability using SVM-RBF among the set of Top K-union ROIs on 8 working memory tasks from 4 frameworks.}
	
    \setlength{\tabcolsep}{4.5pt}
	\small
	\begin{tabular}{c|ccc|ccc|ccc|ccc}
	    \toprule
		\multirow{1}{*}{\diagbox{K}{Models}} & 
		\multicolumn{3}{c|}{Neurosynth} &
		\multicolumn{3}{c|}{MLP-Mixer + BrainNetX} &
		\multicolumn{3}{c|}{ST-GCN + BrainNetX} &
		\multicolumn{3}{c}{STpGCN + BrainNetX}\\
		& BEST & AVG & STD & BEST & AVG & STD & BEST & AVG & STD & BEST & AVG & STD\\
		\midrule
		1 & 46.9 & 41.1 & 3.4 & 33.7 & \underline{30.8}* & 2.2 & 46.0 & 41.9 & 4.3 & 49.5 & \textbf{46.4}* & 2.5\\
		2 & 49.2 & 45.3 & 3.1 & 39.1 & \underline{33.8}* & 3.0 & 47.0 & 43.3 & 2.2 & 54.6 & \textbf{51.4}* & 3.0\\
		3 & 50.5 & 46.4 & 2.6 & 41.8 & \underline{36.7}* & 2.3 & 56.8 & {52.7}* & 2.5 & 66.6 & \textbf{59.4}* & 3.3\\
		4 & 53.5 & 48.4 & 3.6 & 40.1 & \underline{38.1}* & 1.5 & 59.2 & {55.8}* & 2.8 & 65.1 & \textbf{61.3}* & 2.1\\
		5 & 54.5 & 51.7 & 1.6 & 43.2 & \underline{38.9}* & 2.9 & 61.6 & {56.8}* & 3.3 & 64.0 & \textbf{61.6}* & 2.4\\
		6 & 56.9 & 52.8 & 2.8 & 45.8 & \underline{41.0}* & 2.8 & 64.4 & {61.5}* & 1.8 & 67.2 & \textbf{64.0}* & 2.1\\
		7 & 60.3 & 53.6 & 3.9 & 44.5 & \underline{40.6}* & 3.2 & 66.1 & {62.2}* & 2.3 & 70.7 & \textbf{64.8}* & 3.4\\
		8 & 58.3 & 54.6 & 2.8 & 44.5 & \underline{42.2}* & 1.9 & 68.0 & {63.2}* & 2.6 & 67.4 & \textbf{64.2}* & 2.4\\
		9 & 61.6 & 57.2 & 3.3 & 44.5 & \underline{42.2}* & 1.8 & 66.0 & {61.7}* & 3.2 & 69.0 & \textbf{65.0}* & 1.8\\
		10 & 60.8 & 58.4 & 2.5 & 48.0 & \underline{42.8}* & 3.4 & 68.8 & {64.5}* & 2.7 & 71.3 & \textbf{65.5}* & 2.6\\
		\midrule
		$-/\text{ns}/+$ & \multicolumn{3}{c|}{no comparison} &
		\multicolumn{3}{c|}{10 / 0 / 0} &
		\multicolumn{3}{c|}{0 / 2 / 8} & 
		\multicolumn{3}{c}{0 / 0 / 10}\\
		\bottomrule
	\end{tabular}
	\begin{tablenotes}[flushleft]
    \item Note: ``BEST'' and ``AVG'' stand for the best and averaged Macro F1 score ($\%$) obtained from 10 cross validation. ``STD'' stands for the standard deviation. SVM-RBF is set to have 64 max iteration number and one-versus-rest decision policy. The best results are highlighted in bold, and the worst results are marked with underline.
    ``-/ns/+'' indicate the number of tests that the results are ``statistically worse/not significant/significantly better'' than those obtained by Neurosynth, respectively. * for $p<0.05$ on Wilcoxon rank-sum test compared with the results from Neurosynth baselines.
   \end{tablenotes}
	\end{threeparttable}
\end{table*}

\textbf{Quantitative analysis}: It is obvious that the activity of brain regions associated with different tasks should be easily classified by a simple classifier. Therefore, to further examine whether the selected brain regions are responsible for brain decoding, we used an SVM-RBF to evaluate the brain decoding performance of the annotated brain regions. Specifically, we aim to decode the 8 working memory tasks based on the selected brain regions' fMRI signals using SVM-RBF.
The meta-analysis-based state-of-the-art cognitive tasks annotation method, Neurosynth~\cite{yarkoni2011large}, is applied as the baseline. The list of brain regions annotated by Neurosynth, MLP-Mixer, ST-GCN and STpGCN are shown in Supplementary Table~S1, Table~S2, Table~S3 and Table~S4, respectively.

For Neurosynth, top $K$ important brain regions are selected from the fMRI meta-analysis with the keyword ``working memory''. 
For MLP-Mixer, ST-GCN and STpGCN, we select the top $K$ important ROIs of each working memory task, ranked by the well-trained models. To be noted, we excluded the primary visual area in all tests, in order to prevent the model from fully utilizing the visual response rather than the cognitive state-relevant activity for brain decoding. 
The explainability performance is measured by Macro F1 score in Table~\ref{tab:top_k_roi_svm_comparison}. From these results, we summarize
several interesting findings:
\begin{itemize}
    \item Although MLP-Mixer is only $1\%\sim3\%$ inferior to STpGCN in brain decoding (Table~\ref{tab:deco_result}), MLP-Mixer cannot serve as a reliable model to provide explainable results for brain decoding since the important brain regions it annotates cannot be directly used by SVM-RBF for classification (Table~\ref{tab:top_k_roi_svm_comparison}).
    \item When the hyper-parameter $K$ is small (\ie $K \leq 2$), Neurosynth, ST-GCN and STpGCN have equivalent explainability performance, suggesting they are capable of capturing the top 2-union task-relevant brain regions.
    \item When the hyper-parameter $K$ is relatively large (\ie $K \geq 3$), Neurosynth, ST-GCN and STpGCN show significant differences in explainability. STpGCN has the highest explainability to capture a large number of task-relevant brain regions.
    \item Under different $K$ size settings, STpGCN can robustly annotate the brain regions associated with each cognitive task, indicating that the brain regions found by STpGCN are well ranked according to their importance.
\end{itemize}

\subsection{Generalizability of STpGCN}\label{subsec:generalizability}

We have verified the capability of STpGCN on classifying 23 cognitive states. Here we test the generalizability of STpGCN to other types of the classification task, \ie classifying the 4 types of visual stimuli (\ie body, face, place, tool) in 8 working memory (WM) tasks. 
In Table~\ref{tab:wm_classification_result}, we compare STpGCN to the state-of-the-art graph models on the above stimuli classification task. Based on the results, STpGCN outperforms other methods across all four metrics and achieves a decoding performance of 87.8\%. STpGCN performs much better than the graph models without time-dependent processing modules (\ie GCN, GIN and GAT).

\begin{figure*}[htbp]
  \centering
  
  \subfloat{
    \label{fig:subfig:wm} 
     \includegraphics[page=1,width=1.3in]{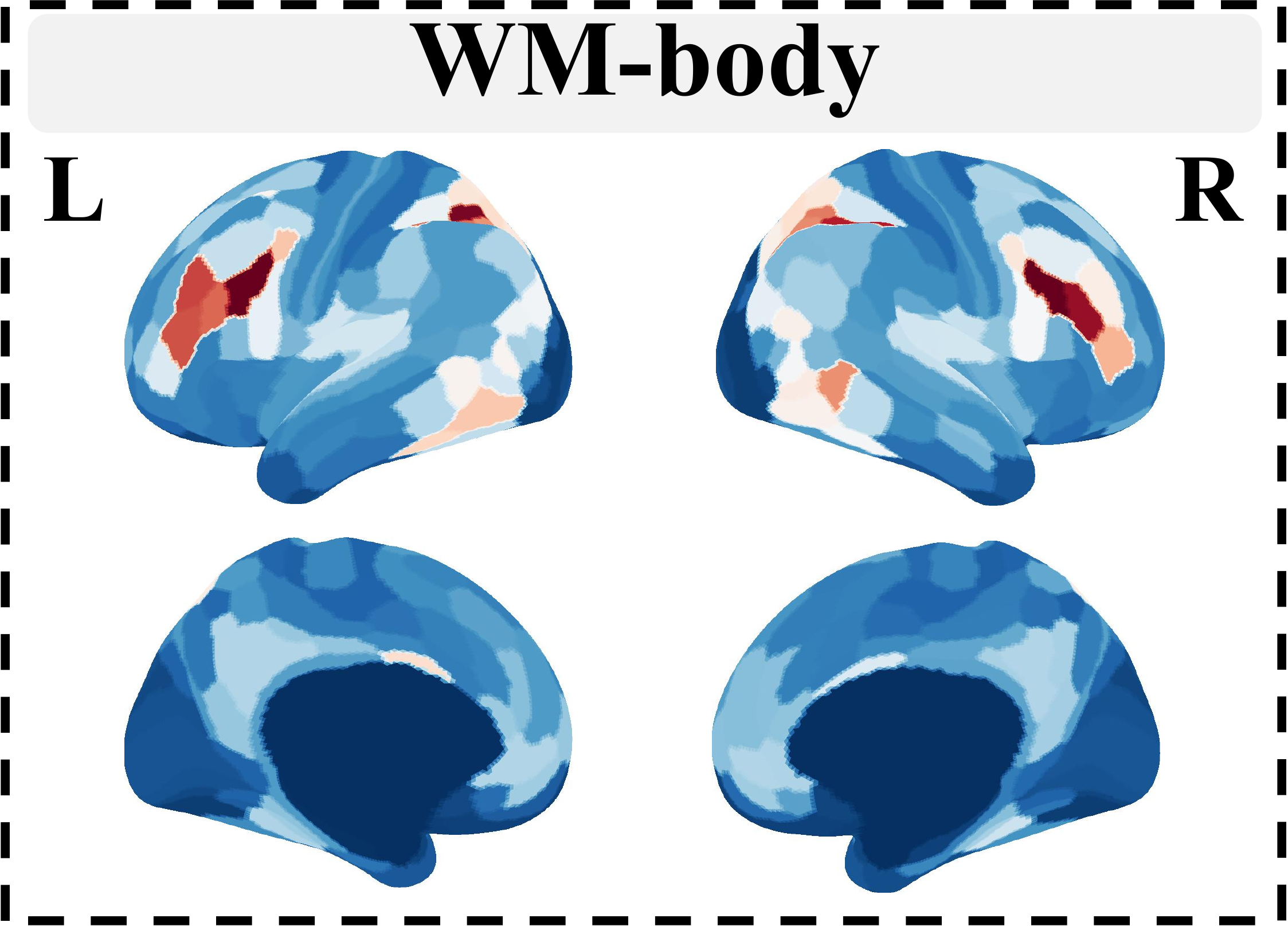}
     }
  \subfloat{
    \label{fig:subfig:mot} 
     \includegraphics[page=2,width=1.3in]{fig/WM-stpgcn.pdf}
     }
  \subfloat{
    \label{fig:subfig:emo} 
    \includegraphics[page=3,width=1.3in]{fig/WM-stpgcn.pdf}
    }
   \subfloat{
    \label{fig:subfig:rel} 
     \includegraphics[page=4,width=1.3in]{fig/WM-stpgcn.pdf}
     }
  \subfloat{
    \label{fig:subfig:bar} 
    \includegraphics[page=1,width=0.2in]{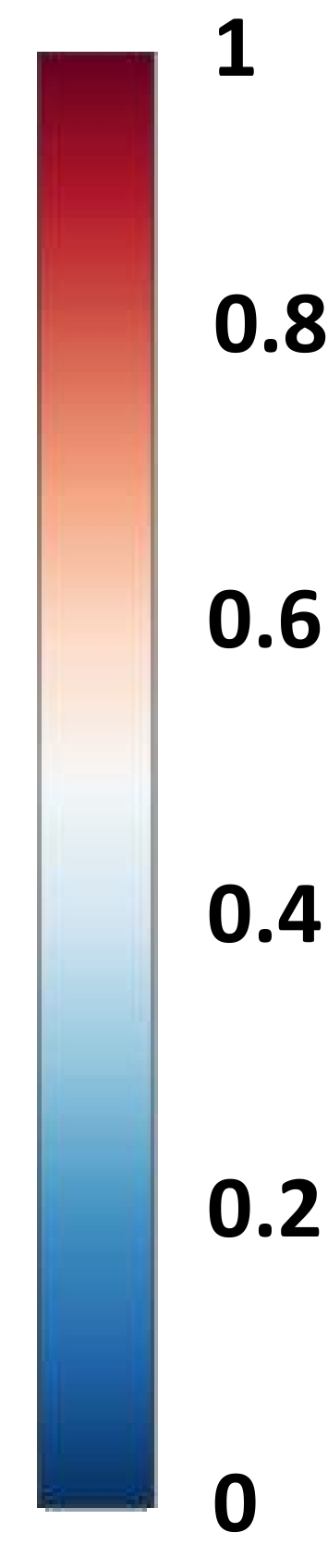}
     }
 
  \caption{Visualization of the averaged importance score of different working memory tasks with the same stimuli by STpGCN and BrainNetX given 15 time points of fMRI with MMP atlas across 10-fold cross-validation.}
  \label{fig:wm_stimuli} 
\end{figure*}

\begin{table}[ht]
    \centering
    \caption{Comparison of the 4-stimuli classification in working memory tasks. We quantify four metrics ($\%$) of STpGCN and some baseline graph models on 15 time points of fMRI with the MMP atlas.}\label{tab:wm_classification_result}
    \small
    \setlength{\tabcolsep}{2pt}
    \begin{tabular}{c|cccc}
        \hline
        \textbf{Methods} & ACC $\uparrow$    & Macro Pre  $\uparrow$ & Macro R$\uparrow$ & Macro F1$\uparrow$ \\ \hline
        GCN & 67.8$\pm$1.2  & 66.0$\pm$4.4     & 66.4$\pm$4.5  &   66.9$\pm$1.4  \\
        GIN & 58.2$\pm$3.4  & 58.8$\pm$2.7     & 58.2$\pm$3.4  &   56.5$\pm$0.4  \\
        GAT & 79.6$\pm$1.5  & 79.9$\pm$1.5     &  79.6$\pm$1.5   &   79.3$\pm$1.5 \\
        ST-GCN & 86.3$\pm$1.3  & 86.3$\pm$1.3     &  86.3$\pm$1.3   &   86.3$\pm$1.3 \\ \hline
        STpGCN (ours) & \textbf{87.8$\pm$1.4}  & \textbf{87.9$\pm$1.4}    & \textbf{87.8$\pm$1.4}    & \textbf{87.8$\pm$1.4}   \\ \hline
    \end{tabular}
    \begin{tablenotes}
     \item Note: The mean $\pm$ std are obtained across 10-fold cross-validation. \\The best results are highlighted in bold.
    \end{tablenotes}
\end{table}

To explain the classification results, we annotate the important brain regions for each task using BrainNetX and STpGCN in Fig.~\ref{fig:wm_stimuli}. To be noted, unlike the brain regions identified in Sec.~\ref{sec:explainability} to distinguish 23 brain states, here we mainly find the brain regions corresponding to different stimuli when subjects are conducting WM task. Our results demonstrate the key brain regions identified by STpGCN and BrainNetX on WM tasks with body and face stimuli are specifically located in the extrastriate body area (EBA) and fusiform face area (FFA), respectively. These two areas are considered to be responsible for processing the body and face images~\cite{bao2020map}. Interestingly, these tasks have a large overlap on the inferior parietal lobule and superior parietal lobule, which have been overlooked by neuroscientists for body and face processes.
The WM task with places stimuli broadly involves multiple brain areas, while the key brain areas for tools stimuli are mainly in the insular and motor areas. The full table of the key brain regions responsible for different working memory tasks with the same stimuli is shown in Supplementary Table~S5.

\section{Discussion}
\label{sec:discussion}

In this study, we demonstrate that STpGCN performs well in decoding 23 brain states based on task-evoked fMRI data. To further explain different data-driven decoding methods, we propose BrainNetX for providing a whole brain explanation from a brain-network perspective. The experimental results conducted on the HCP dataset across different categories of tasks illustrate the explainability, generalizability and robustness characteristics of STpGCN.

\subsection{The brain-inspired model architecture}
Unlike other well-known brain decoding techniques (\eg multivariate pattern analysis (MVPA)~\cite{haxby2001distributed}, RNN-based methods~\cite{li2019interpretable} and CNN-based methods~\cite{wang2020decoding}), GNNs as geometric deep learning models can represent complex topological relationships and interdependencies between data. The state-of-the-art GNN models for information processing (\eg~\cite{yu2018spatio,fang2021ftpg}) choose to apply a single spatial-temporal pathway for the data integration and prediction. However, their network architectures lack biologically structural similarity compared to the brain, which limits their ability to imitate the brain's working mechanism. In this study, we 
raise the ``brain-like network architecture hypothesis'' since the functional network~\cite{huntenburg2018large,buckner2019brain,deco2021revisiting} and anatomical network~\cite{felleman1991distributed,bullmore2009complex,bullmore2012economy} of our brain have been proven to be hierarchical. Recent work based on pyramid network design has achieved great success in the field of time series data processing~\cite{lin2017feature,li2021temporal,wang2021pyramid}, which encourages us to employ this design concept in GNNs to enhance its biological similarity compared to the brain for our decoding task. Following such a design concept, we propose the STpGCN which includes spatial-temporal pathways that effectively identify brain activity from various scales, as well as bottom-up pathways for integrating coarse-grained and fine-grained information.

Not surprisingly, the results in Fig.~\ref{fig:confusion_mat} \& \ref{fig:robust_diff_time_step} and Table~\ref{tab:robust_diff_ratio} demonstrate that STpGCN outperforms other competitive baselines on brain decoding tasks regardless of the fMRI time points and the size of training dataset. Table~\ref{tab:deco_result} further suggests that the fusion of finer-grained spatio-temporal features and coarse-grained spatio-temporal features is important for brain decoding using fMRI time series information.
In particular, the results indicate that the bottom-up information fusion mechanism is even more important than adding extra spatial-temporal pathways. These results explicitly corroborate ``brain-like network architecture hypothesis''.

\subsection{Brain-network-level explainability}
In order to uncover the underlying decoding mechanism behind the deep graph models, explanation techniques are necessary. Some methods have been designed to provide explainability and explanations for GNNs from different aspects~\cite{yuan2020explainability}, including gradient-based methods, decomposition methods, and surrogate methods. Perturbation-based methods~\cite{ying2019gnnexplainer,schlichtkrull2020interpreting,yuan2021explainability} examine output variation in response to varied input perturbations, which implies the relationship between input features with different outputs in an intuitive manner. 
Nevertheless, none of these methods is guaranteed to uncover reasonable and explainable input features from a neuroscience perspective in brain decoding applications. In this work, we ensure that the provided important input features can form a complete brain network (see Section~\ref{subsec:BrainNetX}). Our method offers a model-agnostic tool for sensitivity analysis to explain brain decoding from the brain-network standpoint.

\subsection{Explainable brain decoding}
Decoding brain activity deepens our understanding of the relationship between neural activity and task stimuli, which is fundamental in the neuroscience field. The MVPA plays a vital role in detecting different cognitive states based on voxel-based patterns of fMRI data ~\cite{haxby2001distributed}. However, MVPA hardly captures the high-order, nonlinear relationships between brain activity and task stimuli. Such complex nonlinear relationships can be captured with deep neural networks to help decode brain information. Recently, a CNN-based residual network achieves higher accuracy in the classification of brain task states compared to MVPA~\cite{wang2020decoding}. In contrast to the CNN models in extracting spatial features, the RNN-based models deal with temporal dependencies. Both CNN and RNN have been applied to neural decoding~\cite{li2019interpretable,li2018brain}. To capture long-distance correlations between two brain regions, CNNs require a stack of convolutional kernels to enlarge the receptive field which poses a challenge to the decoding model itself. GNNs have brought new achievements in brain decoding, which have been proven better at decoding neural data with non-Euclidean structure~\cite{li2021braingnn,zhang2021functional}. In this work, we model the brain as a graph and consider modeling the multiple-scale spatial-temporal dynamics of fMRI data.

The decoding results in Fig.~\ref{fig:confusion_mat} indicate that STpGCN can robustly identify the discriminative patterns from different functional signals. Among 7 categories of cognitive tasks, motor tasks, language tasks, social cognition tasks, and emotion processing tasks are the most discernible cognitive states. The annotation visualization by STpGCN (see Fig.~\ref{fig:annotation_stpgcn}) and MLP-Mixer (see Fig.~\ref{fig:annotation_MLP-Mixer}) 
reveal that despite MLP-Mixer being only marginally  inferior to STpGCN in brain decoding, MLP-Mixer (the state-of-the-art deep learning method for Euclidean space data) might be inaccurate in identifying important brain regions and patterns in relation to the task, thus resulting in poor explainability in quantitative analysis(see Table.~\ref{tab:top_k_roi_svm_comparison}). In contrast, STpGCN achieves better performance in brain decoding and explainability analysis. We believe that the reliable explainability of STpGCN might be attributed to two mechanisms in our approach: (1) the pyramid-type spatial-temporal information processing mechanism of STpGCN can capture the various levels of temporal information in neural activity underlying the brain states; (2) the brain-network aspect of the whole-brain explanation mechanism of BrainNetX offers richer information than Neurosynth, whose annotation relies on the meta-analysis of statistical results in previous studies.

\subsection{Limitations and future work}
Even though the current STpGCN has made substantial progress in decoding 23 brain states using a data-driven approach, the following aspects require continued attention. First, the brain graph construction method relies on fixed functional connectivity derived from Pearson correlations between brain regions. Other techniques, such as dynamic functional connectivity~\cite{hutchison2013dynamic,gonzalez2018task,zheng2022kuramoto} might be helpful for constructing dynamic graphs and thus enhancing the robustness of brain decoding performance using dynamic graph networks. Second, the spatial information processing module can be improved by introducing the attention mechanism~\cite{velivckovic2018graph}, which allows brain regions to aggregate weighted information from their neighbors. This may help uncover more fine-grained cues about the separation and integration of different brain regions in the brain when doing different tasks.

\section{Conclusion}\label{sec:conclusion}
The growing tension between the accuracy and explainability of brain decoding has motivated the development of models and the discovery of neural mechanisms under tasks. In this study, we proposed a novel framework, including a graph model (\ie STpGCN) for brain decoding and a brain-network level brain annotation method (\ie BrainNetX) for explainability of brain decoding. STpGCN exploits the multi-scale temporal information and spatial information of the brain graph, resulting in over 90\% averaged decoding accuracy on 23 task-related brain states. BrainNetX can identify the key brain regions related to the task by scoring each brain network on the brain graph. 
Beyond brain decoding, STpGCN holds promise for broad applications, such as brain disease detection~\cite{yuan2021machine}.

\section*{Acknowledgment}
The authors would like to thank Siheng Chen, Pinglei Bao and Shengyuan Cai for their invaluable comments and suggestions that greatly improved this manuscript.

\bibliography{STpGCN-MAIN}

\begin{thebibliography}{10}
\providecommand{\url}[1]{#1}
\csname url@samestyle\endcsname
\providecommand{\newblock}{\relax}
\providecommand{\bibinfo}[2]{#2}
\providecommand{\BIBentrySTDinterwordspacing}{\spaceskip=0pt\relax}
\providecommand{\BIBentryALTinterwordstretchfactor}{4}
\providecommand{\BIBentryALTinterwordspacing}{\spaceskip=\fontdimen2\font plus
\BIBentryALTinterwordstretchfactor\fontdimen3\font minus
  \fontdimen4\font\relax}
\providecommand{\BIBforeignlanguage}[2]{{%
\expandafter\ifx\csname l@#1\endcsname\relax
\typeout{** WARNING: IEEEtran.bst: No hyphenation pattern has been}%
\typeout{** loaded for the language `#1'. Using the pattern for}%
\typeout{** the default language instead.}%
\else
\language=\csname l@#1\endcsname
\fi
#2}}
\providecommand{\BIBdecl}{\relax}
\BIBdecl

\bibitem{deco2015rethinking}
G.~Deco, G.~Tononi, M.~Boly, and M.~L. Kringelbach, ``Rethinking segregation
  and integration: contributions of whole-brain modelling,'' \emph{Nature
  Reviews Neuroscience}, vol.~16, no.~7, pp. 430--439, 2015.

\bibitem{ito2020discovering}
T.~Ito, L.~Hearne, R.~Mill, C.~Cocuzza, and M.~W. Cole, ``Discovering the
  computational relevance of brain network organization,'' \emph{Trends in
  cognitive sciences}, vol.~24, no.~1, pp. 25--38, 2020.

\bibitem{kapogiannis2011disrupted}
D.~Kapogiannis and M.~P. Mattson, ``Disrupted energy metabolism and neuronal
  circuit dysfunction in cognitive impairment and alzheimer's disease,''
  \emph{The Lancet Neurology}, vol.~10, no.~2, pp. 187--198, 2011.

\bibitem{dehaene1998inferring}
S.~Dehaene, G.~Le~Clec'H, L.~Cohen, J.-B. Poline, P.-F. van~de Moortele, and
  D.~Le~Bihan, ``Inferring behavior from functional brain images,''
  \emph{Nature neuroscience}, vol.~1, no.~7, pp. 549--549, 1998.

\bibitem{drobyshevsky2006rapid}
A.~Drobyshevsky, S.~B. Baumann, and W.~Schneider, ``A rapid fmri task battery
  for mapping of visual, motor, cognitive, and emotional function,''
  \emph{Neuroimage}, vol.~31, no.~2, pp. 732--744, 2006.

\bibitem{miller2009unique}
M.~B. Miller, C.-L. Donovan, J.~D. Van~Horn, E.~German, P.~Sokol-Hessner, and
  G.~L. Wolford, ``Unique and persistent individual patterns of brain activity
  across different memory retrieval tasks,'' \emph{Neuroimage}, vol.~48, no.~3,
  pp. 625--635, 2009.

\bibitem{woo2017building}
C.-W. Woo, L.~J. Chang, M.~A. Lindquist, and T.~D. Wager, ``Building better
  biomarkers: brain models in translational neuroimaging,'' \emph{Nature
  neuroscience}, vol.~20, no.~3, pp. 365--377, 2017.

\bibitem{wang2020decoding}
X.~Wang, X.~Liang, Z.~Jiang, B.~A. Nguchu, Y.~Zhou, Y.~Wang, H.~Wang, Y.~Li,
  Y.~Zhu, F.~Wu \emph{et~al.}, ``Decoding and mapping task states of the human
  brain via deep learning,'' \emph{Human brain mapping}, vol.~41, no.~6, pp.
  1505--1519, 2020.

\bibitem{bassett2017network}
D.~S. Bassett and O.~Sporns, ``Network neuroscience,'' \emph{Nature
  neuroscience}, vol.~20, no.~3, pp. 353--364, 2017.

\bibitem{rosenbaum2017spatial}
R.~Rosenbaum, M.~A. Smith, A.~Kohn, J.~E. Rubin, and B.~Doiron, ``The spatial
  structure of correlated neuronal variability,'' \emph{Nature neuroscience},
  vol.~20, no.~1, pp. 107--114, 2017.

\bibitem{kipf2016semi}
T.~N. Kipf and M.~Welling, ``Semi-supervised classification with graph
  convolutional networks,'' in \emph{5th International Conference on Learning
  Representations, ICLR}, 2017.

\bibitem{velivckovic2018graph}
P.~Veli{\v{c}}kovi{\'c}, G.~Cucurull, A.~Casanova, A.~Romero, P.~Li{\`o}, and
  Y.~Bengio, ``Graph attention networks,'' in \emph{6th International
  Conference on Learning Representations, ICLR}, 2018.

\bibitem{li2021braingnn}
X.~Li, Y.~Zhou, N.~Dvornek, M.~Zhang, S.~Gao, J.~Zhuang, D.~Scheinost, L.~H.
  Staib, P.~Ventola, and J.~S. Duncan, ``Braingnn: Interpretable brain graph
  neural network for fmri analysis,'' \emph{Medical Image Analysis}, vol.~74,
  p. 102233, 2021.

\bibitem{zhang2021functional}
Y.~Zhang, L.~Tetrel, B.~Thirion, and P.~Bellec, ``Functional annotation of
  human cognitive states using deep graph convolution,'' \emph{NeuroImage},
  vol. 231, p. 117847, 2021.

\bibitem{yu2018spatio}
B.~Yu, H.~Yin, and Z.~Zhu, ``Spatio-temporal graph convolutional networks: a
  deep learning framework for traffic forecasting,'' in \emph{Proceedings of
  the 27th International Joint Conference on Artificial Intelligence}, 2018,
  pp. 3634--3640.

\bibitem{bullmore2009complex}
E.~Bullmore and O.~Sporns, ``Complex brain networks: graph theoretical analysis
  of structural and functional systems,'' \emph{Nature reviews neuroscience},
  vol.~10, no.~3, pp. 186--198, 2009.

\bibitem{bullmore2012economy}
{Bullmore, Ed and Sporns, Olaf}, ``The economy of brain network organization,''
  \emph{Nature reviews neuroscience}, vol.~13, no.~5, pp. 336--349, 2012.

\bibitem{buckner2019brain}
R.~L. Buckner and L.~M. DiNicola, ``The brain’s default network: updated
  anatomy, physiology and evolving insights,'' \emph{Nature Reviews
  Neuroscience}, vol.~20, no.~10, pp. 593--608, 2019.

\bibitem{deco2021revisiting}
G.~Deco, D.~Vidaurre, and M.~L. Kringelbach, ``Revisiting the global workspace
  orchestrating the hierarchical organization of the human brain,''
  \emph{Nature human behaviour}, vol.~5, no.~4, pp. 497--511, 2021.

\bibitem{ding2016cortical}
N.~Ding, L.~Melloni, H.~Zhang, X.~Tian, and D.~Poeppel, ``Cortical tracking of
  hierarchical linguistic structures in connected speech,'' \emph{Nature
  neuroscience}, vol.~19, no.~1, pp. 158--164, 2016.

\bibitem{lin2017feature}
T.-Y. Lin, P.~Doll{\'a}r, R.~Girshick, K.~He, B.~Hariharan, and S.~Belongie,
  ``Feature pyramid networks for object detection,'' in \emph{Proceedings of
  the IEEE conference on computer vision and pattern recognition}, 2017, pp.
  2117--2125.

\bibitem{van2013wu}
D.~C. Van~Essen, S.~M. Smith, D.~M. Barch, T.~E. Behrens, E.~Yacoub,
  K.~Ugurbil, W.-M.~H. Consortium \emph{et~al.}, ``The wu-minn human connectome
  project: an overview,'' \emph{Neuroimage}, vol.~80, pp. 62--79, 2013.

\bibitem{yarkoni2011large}
T.~Yarkoni, R.~A. Poldrack, T.~E. Nichols, D.~C. Van~Essen, and T.~D. Wager,
  ``Large-scale automated synthesis of human functional neuroimaging data,''
  \emph{Nature methods}, vol.~8, no.~8, pp. 665--670, 2011.

\bibitem{van2009functionally}
M.~P. Van Den~Heuvel, R.~C. Mandl, R.~S. Kahn, and H.~E. Hulshoff~Pol,
  ``Functionally linked resting-state networks reflect the underlying
  structural connectivity architecture of the human brain,'' \emph{Human brain
  mapping}, vol.~30, no.~10, pp. 3127--3141, 2009.

\bibitem{hermundstad2013structural}
A.~M. Hermundstad, D.~S. Bassett, K.~S. Brown, E.~M. Aminoff, D.~Clewett,
  S.~Freeman, A.~Frithsen, A.~Johnson, C.~M. Tipper, M.~B. Miller
  \emph{et~al.}, ``Structural foundations of resting-state and task-based
  functional connectivity in the human brain,'' \emph{Proceedings of the
  National Academy of Sciences}, vol. 110, no.~15, pp. 6169--6174, 2013.

\bibitem{hammond2011wavelets}
D.~K. Hammond, P.~Vandergheynst, and R.~Gribonval, ``Wavelets on graphs via
  spectral graph theory,'' \emph{Applied and Computational Harmonic Analysis},
  vol.~30, no.~2, pp. 129--150, 2011.

\bibitem{defferrard2016convolutional}
M.~Defferrard, X.~Bresson, and P.~Vandergheynst, ``Convolutional neural
  networks on graphs with fast localized spectral filtering,'' \emph{Advances
  in neural information processing systems}, vol.~29, pp. 3844--3852, 2016.

\bibitem{zhang2019view}
P.~Zhang, C.~Lan, J.~Xing, W.~Zeng, J.~Xue, and N.~Zheng, ``View adaptive
  neural networks for high performance skeleton-based human action
  recognition,'' \emph{IEEE transactions on pattern analysis and machine
  intelligence}, vol.~41, no.~8, pp. 1963--1978, 2019.

\bibitem{zhang2020neural}
B.~Zhang, D.~Xiong, J.~Xie, and J.~Su, ``Neural machine translation with
  gru-gated attention model,'' \emph{IEEE transactions on neural networks and
  learning systems}, vol.~31, no.~11, pp. 4688--4698, 2020.

\bibitem{schlichtkrull2021interpreting}
M.~S. Schlichtkrull, N.~De~Cao, and I.~Titov, ``Interpreting graph neural
  networks for nlp with differentiable edge masking,'' in \emph{9th
  International Conference on Learning Representations, ICLR}, 2021.

\bibitem{melgani2004classification}
F.~Melgani and L.~Bruzzone, ``Classification of hyperspectral remote sensing
  images with support vector machines,'' \emph{IEEE Transactions on geoscience
  and remote sensing}, vol.~42, no.~8, pp. 1778--1790, 2004.

\bibitem{tolstikhin2021mlp}
I.~O. Tolstikhin, N.~Houlsby, A.~Kolesnikov, L.~Beyer, X.~Zhai, T.~Unterthiner,
  J.~Yung, A.~Steiner, D.~Keysers, J.~Uszkoreit \emph{et~al.}, ``Mlp-mixer: An
  all-mlp architecture for vision,'' \emph{Advances in Neural Information
  Processing Systems}, vol.~34, 2021.

\bibitem{xu2018powerful}
K.~Xu, W.~Hu, J.~Leskovec, and S.~Jegelka, ``How powerful are graph neural
  networks?'' in \emph{International Conference on Learning Representations},
  2018.

\bibitem{tzourio2002automated}
N.~Tzourio-Mazoyer, B.~Landeau, D.~Papathanassiou, F.~Crivello, O.~Etard,
  N.~Delcroix, B.~Mazoyer, and M.~Joliot, ``Automated anatomical labeling of
  activations in spm using a macroscopic anatomical parcellation of the mni mri
  single-subject brain,'' \emph{Neuroimage}, vol.~15, no.~1, pp. 273--289,
  2002.

\bibitem{glasser2016multi}
M.~F. Glasser, T.~S. Coalson, E.~C. Robinson, C.~D. Hacker, J.~Harwell,
  E.~Yacoub, K.~Ugurbil, J.~Andersson, C.~F. Beckmann, M.~Jenkinson
  \emph{et~al.}, ``A multi-modal parcellation of human cerebral cortex,''
  \emph{Nature}, vol. 536, no. 7615, pp. 171--178, 2016.

\bibitem{thomas2011organization}
B.~Thomas~Yeo, F.~M. Krienen, J.~Sepulcre, M.~R. Sabuncu, D.~Lashkari,
  M.~Hollinshead, J.~L. Roffman, J.~W. Smoller, L.~Z{\"o}llei, J.~R. Polimeni
  \emph{et~al.}, ``The organization of the human cerebral cortex estimated by
  intrinsic functional connectivity,'' \emph{Journal of neurophysiology}, vol.
  106, no.~3, pp. 1125--1165, 2011.

\bibitem{liu2017detecting}
Q.~Liu, S.~Farahibozorg, C.~Porcaro, N.~Wenderoth, and D.~Mantini, ``Detecting
  large-scale networks in the human brain using high-density
  electroencephalography,'' \emph{Human brain mapping}, vol.~38, no.~9, pp.
  4631--4643, 2017.

\bibitem{preti2019decoupling}
M.~G. Preti and D.~Van De~Ville, ``Decoupling of brain function from structure
  reveals regional behavioral specialization in humans,'' \emph{Nature
  communications}, vol.~10, no.~1, pp. 1--7, 2019.

\bibitem{zheng2021graph}
S.~Zheng, D.~Punia, H.~Wu, and Q.~Liu, ``Graph theoretic analysis reveals
  intranasal oxytocin induced network changes over frontal regions,''
  \emph{Neuroscience}, vol. 459, pp. 153--165, 2021.

\bibitem{curtis2003persistent}
C.~E. Curtis and M.~D'Esposito, ``Persistent activity in the prefrontal cortex
  during working memory,'' \emph{Trends in cognitive sciences}, vol.~7, no.~9,
  pp. 415--423, 2003.

\bibitem{bao2020map}
P.~Bao, L.~She, M.~McGill, and D.~Y. Tsao, ``A map of object space in primate
  inferotemporal cortex,'' \emph{Nature}, vol. 583, no. 7814, pp. 103--108,
  2020.

\bibitem{haxby2001distributed}
J.~V. Haxby, M.~I. Gobbini, M.~L. Furey, A.~Ishai, J.~L. Schouten, and
  P.~Pietrini, ``Distributed and overlapping representations of faces and
  objects in ventral temporal cortex,'' \emph{Science}, vol. 293, no. 5539, pp.
  2425--2430, 2001.

\bibitem{li2019interpretable}
H.~Li and Y.~Fan, ``Interpretable, highly accurate brain decoding of subtly
  distinct brain states from functional mri using intrinsic functional networks
  and long short-term memory recurrent neural networks,'' \emph{NeuroImage},
  vol. 202, p. 116059, 2019.

\bibitem{fang2021ftpg}
M.~Fang, L.~Tang, X.~Yang, Y.~Chen, C.~Li, and Q.~Li, ``Ftpg: A fine-grained
  traffic prediction method with graph attention network using big trace
  data,'' \emph{IEEE Transactions on Intelligent Transportation Systems}, 2021.

\bibitem{huntenburg2018large}
J.~M. Huntenburg, P.-L. Bazin, and D.~S. Margulies, ``Large-scale gradients in
  human cortical organization,'' \emph{Trends in cognitive sciences}, vol.~22,
  no.~1, pp. 21--31, 2018.

\bibitem{felleman1991distributed}
D.~J. Felleman and D.~C. Van~Essen, ``Distributed hierarchical processing in
  the primate cerebral cortex.'' \emph{Cerebral cortex (New York, NY: 1991)},
  vol.~1, no.~1, pp. 1--47, 1991.

\bibitem{li2021temporal}
Y.~Li, R.~Liang, W.~Wei, W.~Wang, J.~Zhou, and X.~Li, ``Temporal pyramid
  network with spatial-temporal attention for pedestrian trajectory
  prediction,'' \emph{IEEE Transactions on Network Science and Engineering},
  2021.

\bibitem{wang2021pyramid}
Y.~Wang, P.~Zhang, S.~Gao, X.~Geng, H.~Lu, and D.~Wang, ``Pyramid
  spatial-temporal aggregation for video-based person re-identification,'' in
  \emph{Proceedings of the IEEE/CVF International Conference on Computer
  Vision}, 2021, pp. 12\,026--12\,035.

\bibitem{yuan2020explainability}
H.~Yuan, H.~Yu, S.~Gui, and S.~Ji, ``Explainability in graph neural networks: A
  taxonomic survey,'' \emph{arXiv preprint arXiv:2012.15445}, 2020.

\bibitem{ying2019gnnexplainer}
Z.~Ying, D.~Bourgeois, J.~You, M.~Zitnik, and J.~Leskovec, ``Gnnexplainer:
  Generating explanations for graph neural networks,'' \emph{Advances in neural
  information processing systems}, vol.~32, 2019.

\bibitem{schlichtkrull2020interpreting}
M.~S. Schlichtkrull, N.~De~Cao, and I.~Titov, ``Interpreting graph neural
  networks for nlp with differentiable edge masking,'' in \emph{International
  Conference on Learning Representations}, 2020.

\bibitem{yuan2021explainability}
H.~Yuan, H.~Yu, J.~Wang, K.~Li, and S.~Ji, ``On explainability of graph neural
  networks via subgraph explorations,'' in \emph{International Conference on
  Machine Learning}.\hskip 1em plus 0.5em minus 0.4em\relax PMLR, 2021, pp.
  12\,241--12\,252.

\bibitem{li2018brain}
{Li, Hongming and Fan, Yong}, ``Brain decoding from functional mri using long
  short-term memory recurrent neural networks,'' in \emph{International
  Conference on Medical Image Computing and Computer-Assisted
  Intervention}.\hskip 1em plus 0.5em minus 0.4em\relax Springer, 2018, pp.
  320--328.

\bibitem{hutchison2013dynamic}
R.~M. Hutchison, T.~Womelsdorf, E.~A. Allen, P.~A. Bandettini, V.~D. Calhoun,
  M.~Corbetta, S.~Della~Penna, J.~H. Duyn, G.~H. Glover, J.~Gonzalez-Castillo
  \emph{et~al.}, ``Dynamic functional connectivity: promise, issues, and
  interpretations,'' \emph{Neuroimage}, vol.~80, pp. 360--378, 2013.

\bibitem{gonzalez2018task}
J.~Gonzalez-Castillo and P.~A. Bandettini, ``Task-based dynamic functional
  connectivity: Recent findings and open questions,'' \emph{Neuroimage}, vol.
  180, pp. 526--533, 2018.

\bibitem{zheng2022kuramoto}
S.~Zheng, Z.~Liang, Y.~Qu, Q.~Wu, H.~Wu, and Q.~Liu, ``Kuramoto model-based
  analysis reveals oxytocin effects on brain network dynamics,''
  \emph{International Journal of Neural Systems}, vol.~32, no.~02, p. 2250002,
  2022.

\bibitem{yuan2021machine}
J.~Yuan, X.~Ran, K.~Liu, C.~Yao, Y.~Yao, H.~Wu, and Q.~Liu, ``Machine learning
  applications on neuroimaging for diagnosis and prognosis of epilepsy: A
  review,'' \emph{Journal of neuroscience methods}, p. 109441, 2021.

\bibitem{ye2022spatial}
Z.~Ye, Y.~Qu, Z.~Liang, M.~Wang, and Q.~Liu, ``Explainable fmri-based brain
  decoding via spatial temporal-pyramid graph convolutional network,''
  \emph{Under review}, 2022.

\end{thebibliography}

\newpage
\begin{center}
\textbf{\LARGE Supplementary Materials for ``Explainable fMRI-based Brain Decoding via Spatial Temporal-pyramid Graph Convolutional Network''}
\end{center}

In this paper, we formulated the brain decoding as a task-related brain states classification problem. To tackle the problem, we proposed a spatial-temporal pyramid graph convolutional networks (STpGCN) for learning the representations of neural activities. To examine the interpretability of STpGCN, BrainNetX has been proposed to interpret the decoding results.

In the main manuscript, our approaches compared to some competing baselines. Among all baselines, MLP-Mixer and the state-of-the-art spatio-temporal graph convolutional networks (ST-GCN)~\cite{yu2018spatio} provide highly accurate brain decoding performance. Nevertheless, their performance on explainability varies greatly.

To make the main manuscript more compact, we provide the supplementary materials and organize them as follows. The brain decoding results with AAL atlas are detailed in Section~\ref{sec:deco_aal}. In Section~\ref{sec:variants_stpgcn}, we visualize the architectures of different STpGCN variants. Section~\ref{sec:brainlike} verifies our brain-like network hypothesis. Section~\ref{sec:neux} provides the details of BrainNetX. Section~\ref{sec:add_annotation} illustrates the brain annotation results by competing methods. The details setting of the quantitative explainability experiments are presented in Section~\ref{sec:neurosynth}. The specific brain regions of different methods on distinct tasks are given in Section~\ref{sec:annotation_table}.

\section{Brain Decoding Results with AAL Atlas}~\label{sec:deco_aal}

In this section, we further describe the performance of STpGCN in brain decoding under different atlas.

\begin{figure*}[htbp]
\centering
\includegraphics[width=5.3in]{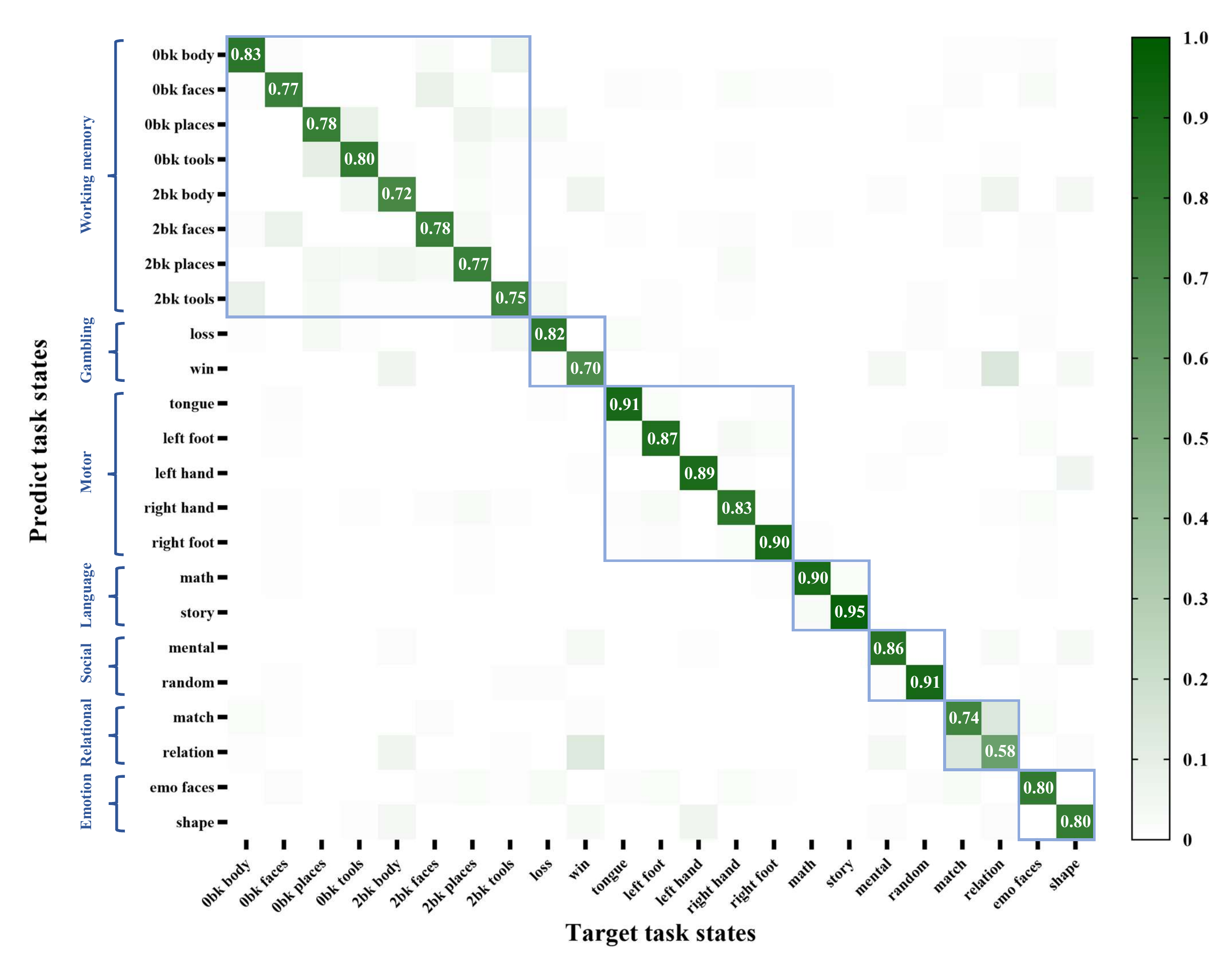}
\caption{Confusion matrix of 23 task-related brain states classification on 15 time steps of fMRI with AAL atlas using STpGCN.
} \label{fig:con_mat_aal}
\end{figure*}

The confusion matrix of 23 brain states prediction by STpGCN using AAL atlas~\cite{tzourio2002automated} (Fig.~\ref{fig:con_mat_aal}) produces a block diagonal which is similar to the result using MML atlas in the main manuscript. Based on the contrast between the confusion matrix from the AAL atlas and the MMP atlas, we can suggest that finer brain segmentation will lead to improved accuracy, especially for higher-order cognitive tasks (\eg relational processing tasks, emotion processing tasks and gambling tasks).


\begin{figure*}[htbp]
  \centering
  \subfloat{
    \label{fig:stpgcn_alpha} 
     \includegraphics[width=5.3in]{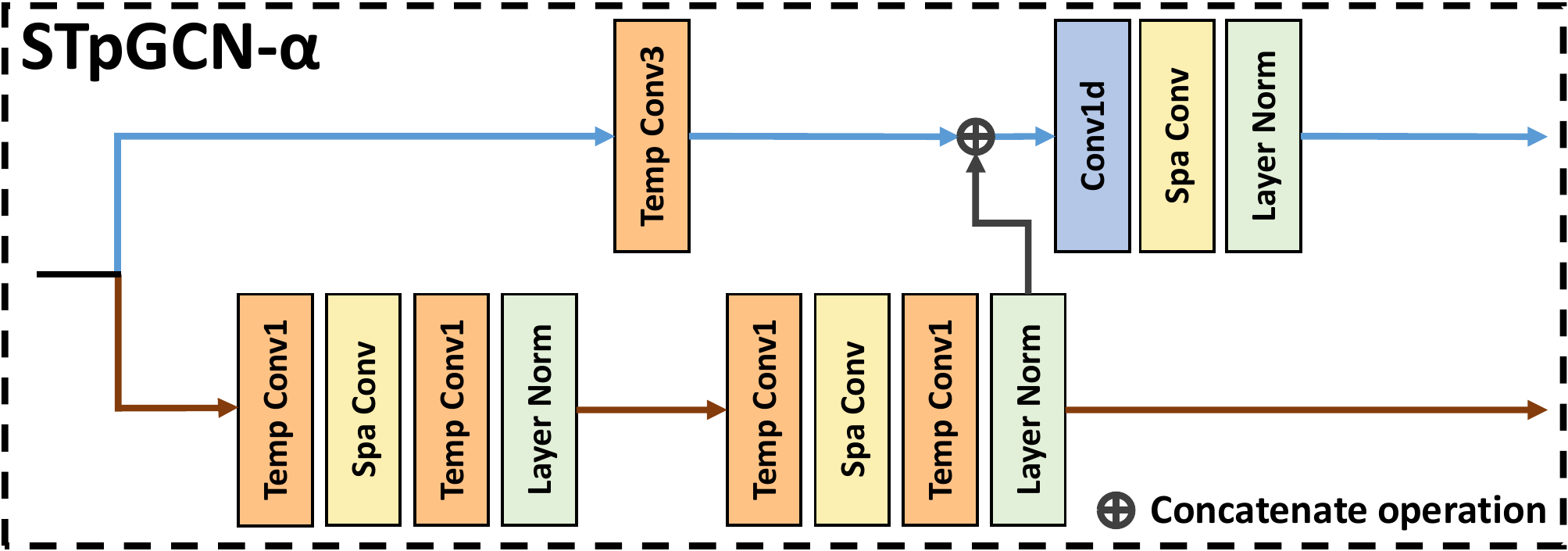}
     } \quad
     
  \subfloat{
    \label{fig:stpgcn_beta} 
     \includegraphics[width=5.3in]{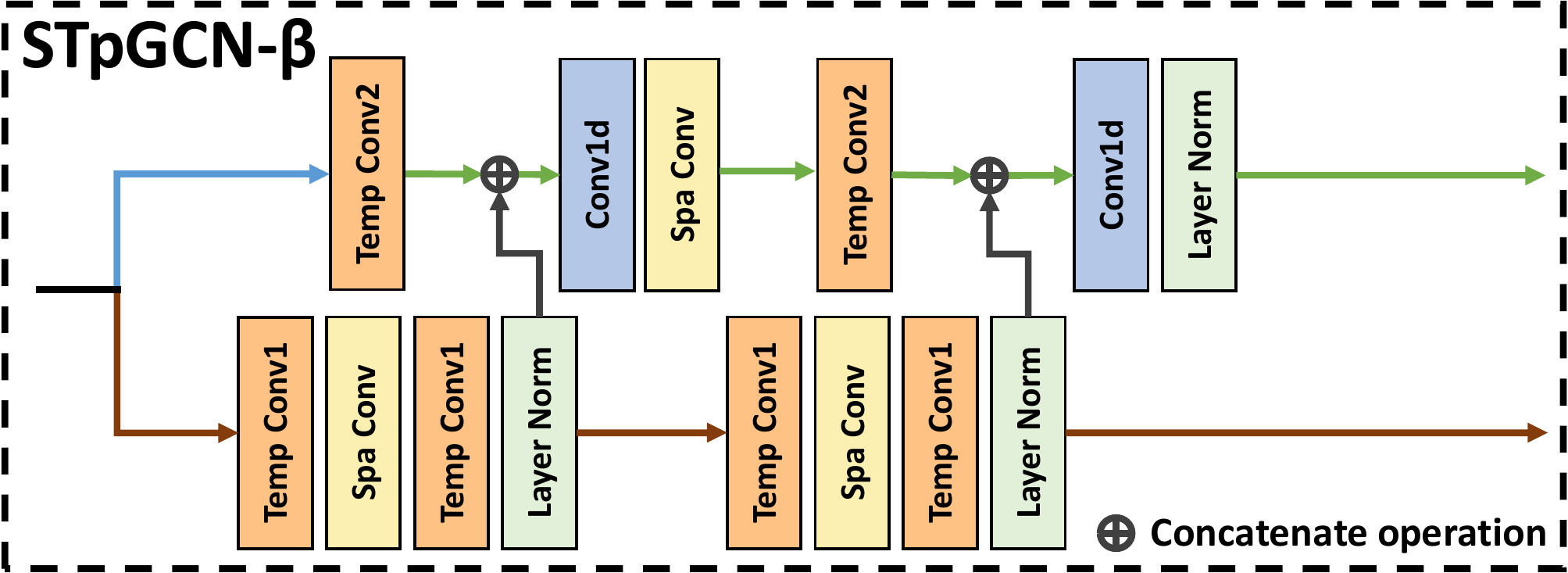}
     } \quad
     
  \subfloat{
    \label{fig:stpgcn_gamma} 
    \includegraphics[width=5.3in]{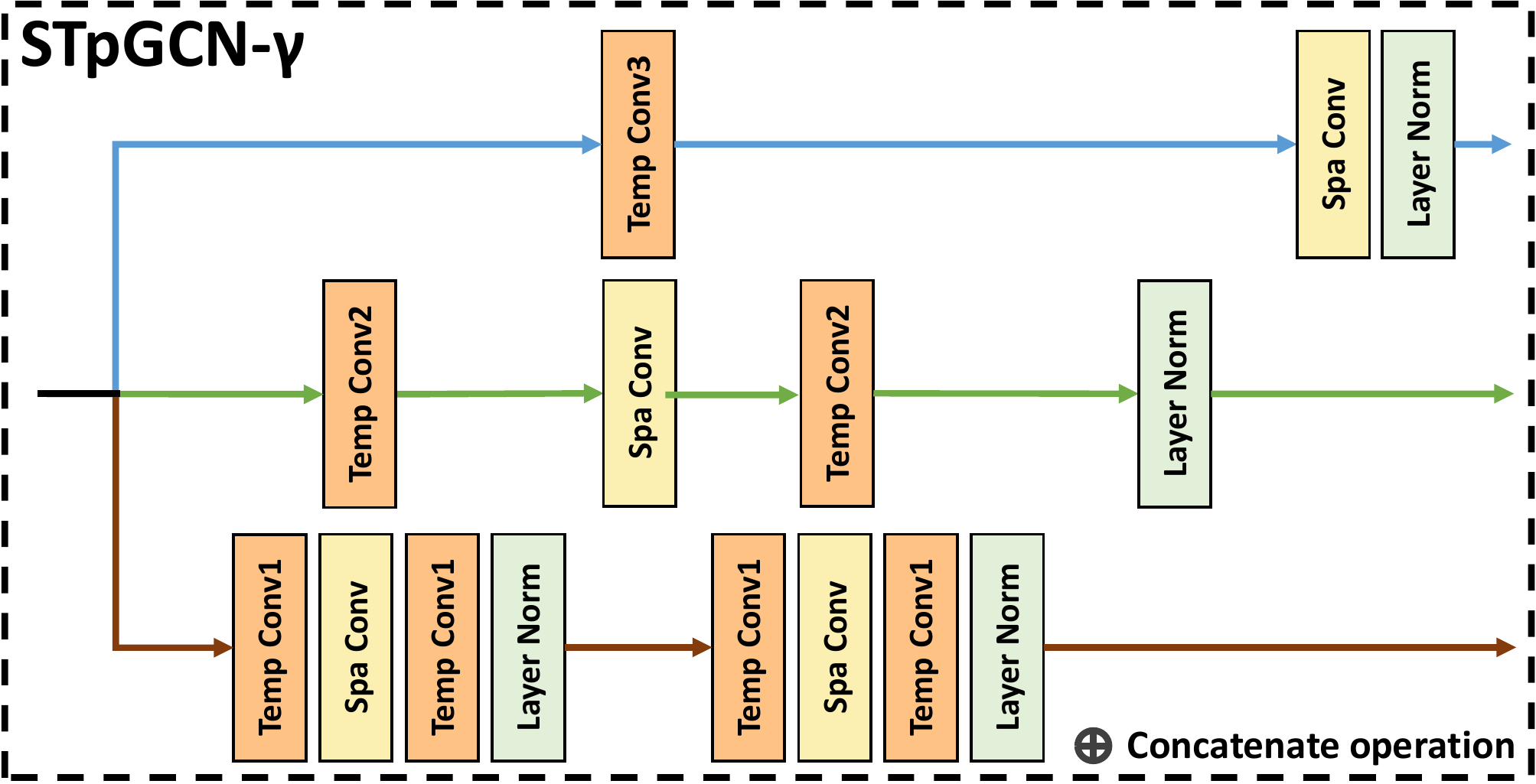}
    } \quad
  \caption{The network architecture used in the ablation study: (a) STpGCN-$\alpha$, where we remove the original middle spatial-temporal pathway in original STpGCN, (b) STpGCN-$\beta$, where we remove the original top spatial-temporal pathway in original STpGCN, (c) STpGCN-$\gamma$, where we remove the original bottom-up pathway in original STpGCN.}
  \label{fig:ablation} 
\end{figure*}

\section{Variants of STpGCN}~\label{sec:variants_stpgcn}
For a better understanding of STpGCN with different modules removed in the ablation study of the main manuscript, in the section below we show the visualization of the architecture of these variants. 

The architectures of STpGCN-$\alpha$, STpGCN-$\beta$ and STpGCN-$\gamma$ are shown in Fig.~\ref{fig:stpgcn_alpha}, Fig.~\ref{fig:stpgcn_beta} and Fig.~\ref{fig:stpgcn_gamma}, respectively.
The STpGCN-$\alpha$ is a variants of STpGCN without the middle spatial-temporal pathway. The STpGCN-$\beta$ is a variants of STpGCN without the top spatial-temporal pathway. The STpGCN-$\gamma$ is a variants of STpGCN without the bottom-up pathway.


\begin{figure*}[htbp]
\centering
\includegraphics[width=5.3in]{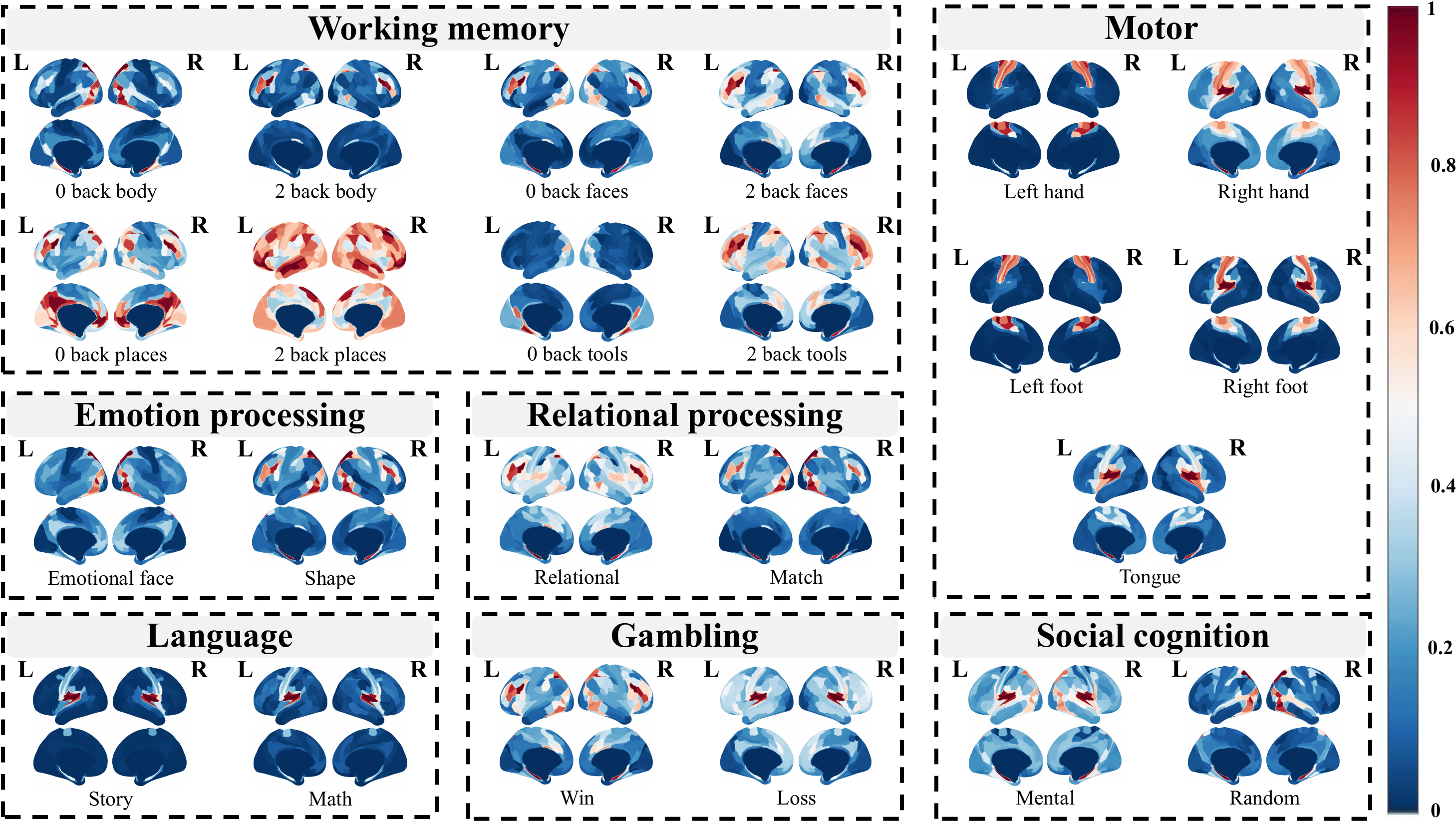}
\caption{Visualization of the importance score of 23 task-related brain states using ST-GCN and BrainNetX given 15 time steps of fMRI with MMP atlas.
} \label{fig:annotation_stgcn}
\end{figure*}

\begin{figure*}[ht]
  \centering
  \subfloat{\label{fig:subfig:feats_ori1}
    \includegraphics[page=1,width=2.6in]{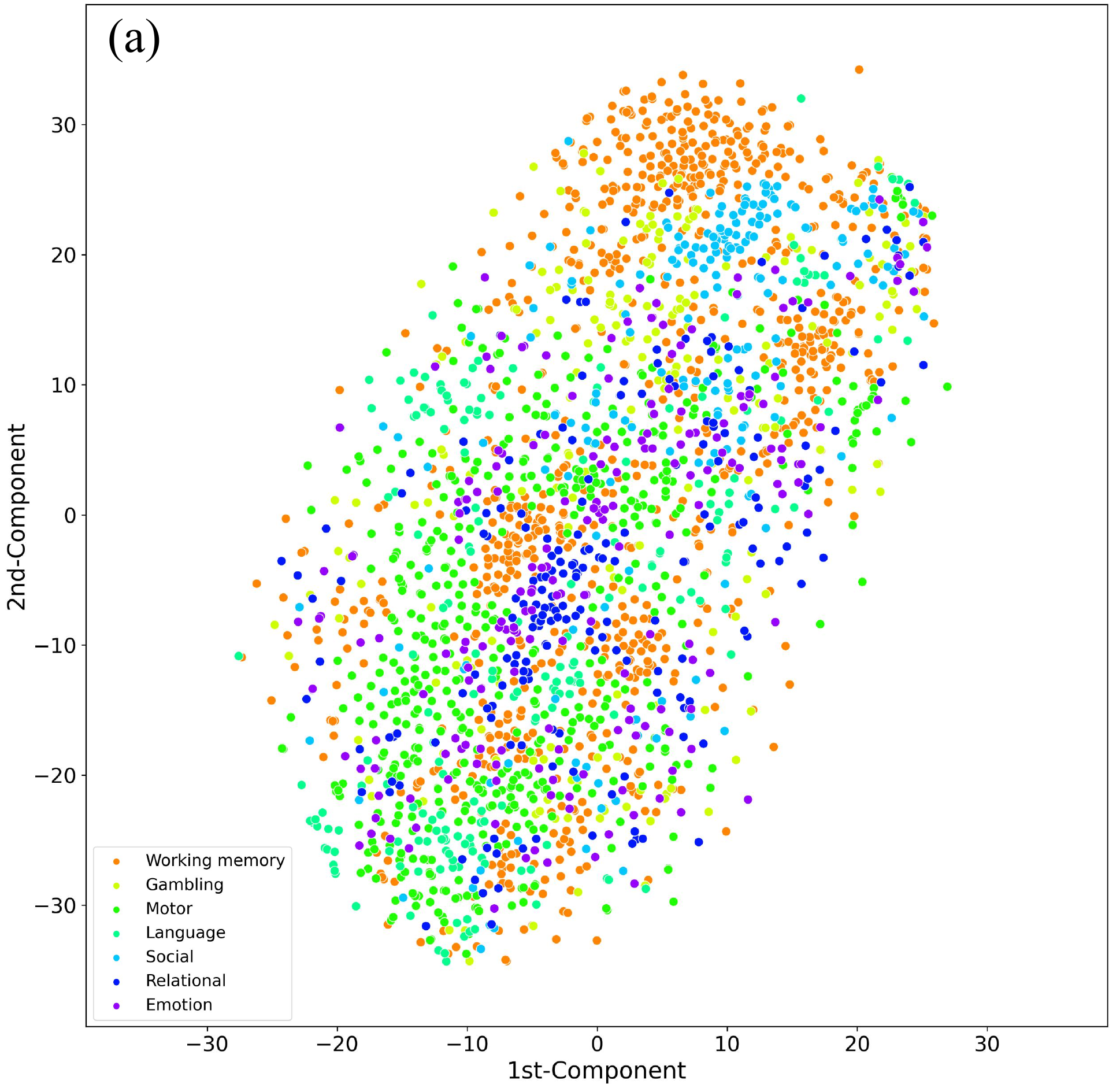}}
  \subfloat{\label{fig:subfig:feats_ori2}
     \includegraphics[page=2,width=2.6in]{fig/tsne/tsne_ori.pdf}}
     
 \subfloat{\label{fig:subfig:feats_stpgcn1}
     \includegraphics[page=1,width=2.6in]{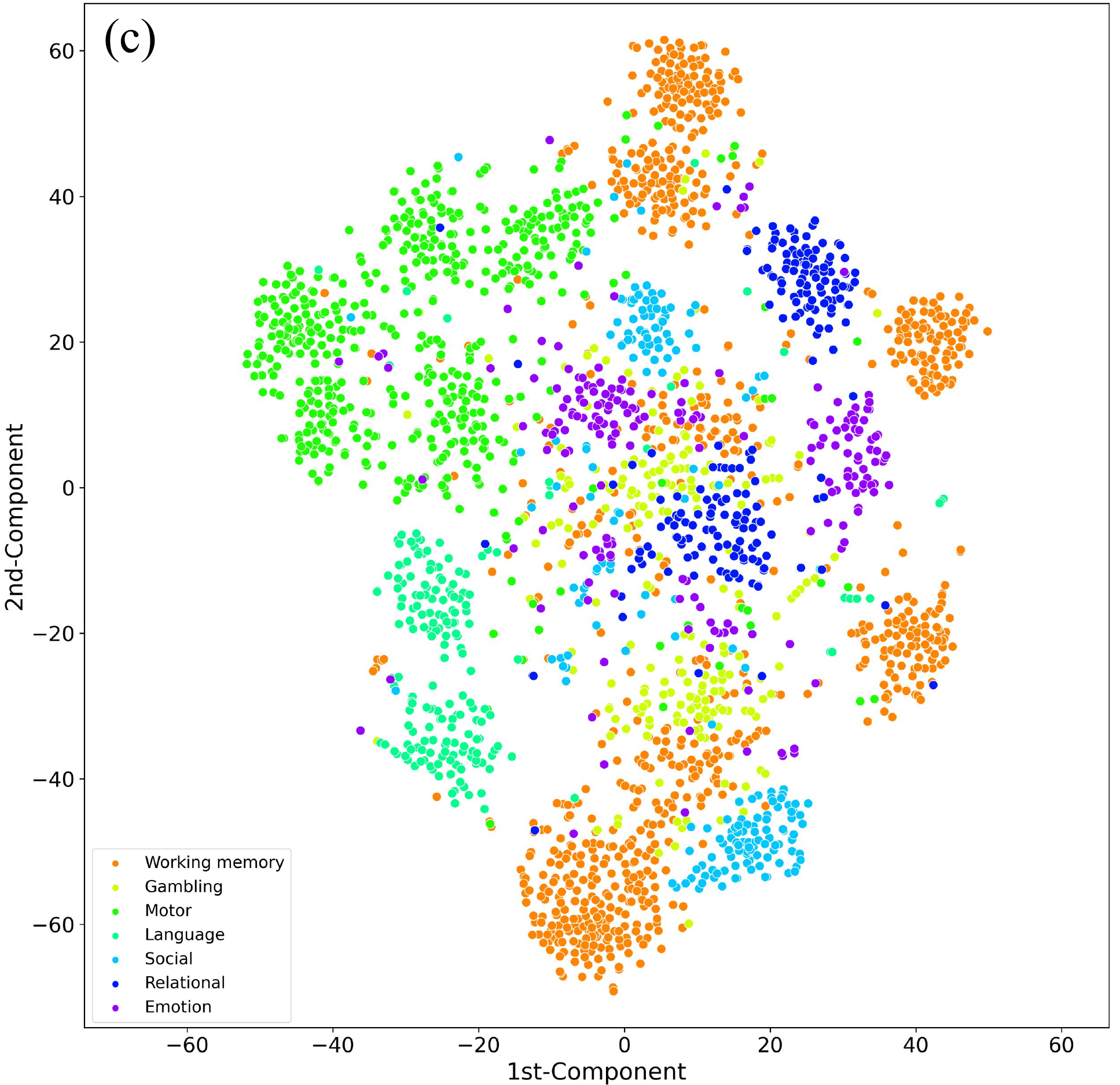}}
 \subfloat{\label{fig:subfig:feats_stpgcn2}
     \includegraphics[page=2,width=2.6in]{fig/tsne/tsne_stpgcn.pdf}}
  \caption{Visualization of the feature space on $15$ time steps fMRI with MMP atlas using t-SNE. We label a tSNE result in two ways, namely as $7$ cognitive categories (left) and as $23$ cognitive tasks (right) where \textbf{(a, b)} denote tSNE results of original input and \textbf{(c, d)} indicate tSNE results of the embedded features after STpGCN.}
  \label{fig:tsne} 
\end{figure*}

\begin{figure*}[htbp]
  \centering
  \subfloat{
    \label{fig:subfig:wm} 
     \includegraphics[page=7,width=1.25in]{fig/supplementary/groundtruth.pdf}
     }
  \subfloat{
    \label{fig:subfig:mot} 
     \includegraphics[page=4,width=1.25in]{fig/supplementary/groundtruth.pdf}
     }
  \subfloat{
    \label{fig:subfig:emo} 
    \includegraphics[page=1,width=1.25in]{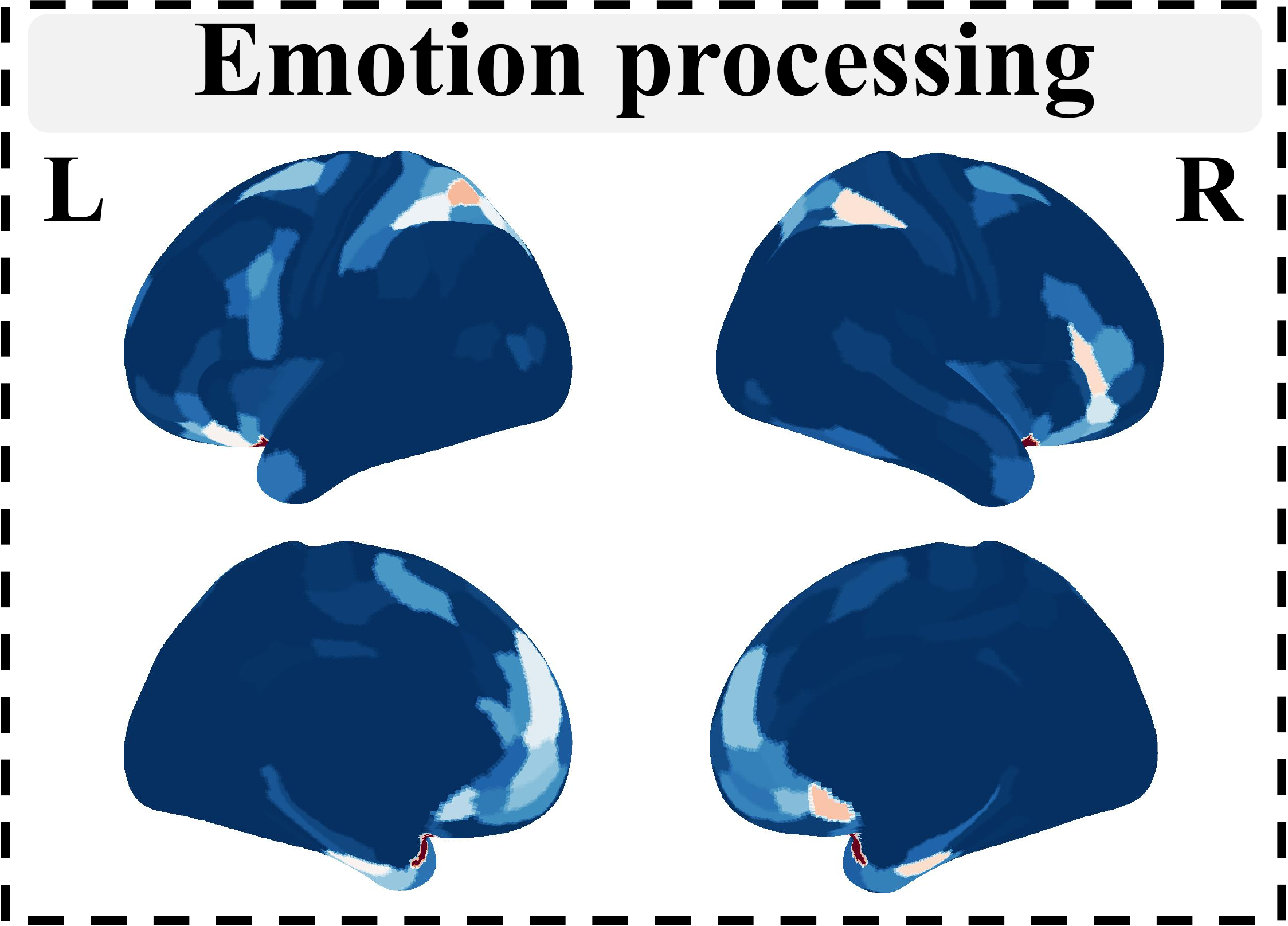}
    }
  \subfloat{
    \label{fig:subfig:bar} 
    \includegraphics[page=1,width=0.2in]{fig/supplementary/bar.pdf}
     }
  
  \subfloat{
    \label{fig:subfig:rel} 
     \includegraphics[page=6,width=1.25in]{fig/supplementary/groundtruth.pdf}
     }
  \subfloat{
    \label{fig:subfig:lan} 
    \includegraphics[page=3,width=1.25in]{fig/supplementary/groundtruth.pdf}
     }
  \subfloat{
    \label{fig:subfig:gam} 
     \includegraphics[page=2,width=1.25in]{fig/supplementary/groundtruth.pdf}
    }
  \subfloat{
    \label{fig:subfig:soc} 
     \includegraphics[page=5,width=1.25in]{fig/supplementary/groundtruth.pdf}
    }
 
  \caption{The visualization of the importance score transformed from the statistical results of \textit{Neurosynth}~\cite{yarkoni2011large} which utilizes term-based meta-analyses from numerous published neuroimaging articles.}
  \label{fig:neurosynth} 
\end{figure*}

\section{Verification of Brain-like Network Architecture Hypothesis}\label{sec:brainlike}
We compare the t-distributed stochastic neighborhood embedding (t-SNE) visualization of the feature space of the original features before STpGCN and the embedded features after STpGCN (Fig.~\ref{fig:tsne}). The results show that the original features cannot separate 7 categories neither 23 tasks (Fig.~\ref{fig:tsne} (a,b)), but the features extracted by STpGCN~(Fig.~\ref{fig:tsne} (a,b)) can, suggesting the benefits of STpGCN for brain decoding. Except for the working memory tasks, the features of the tasks from the same category extracted by STpGCN are rather focal and well clustered.

The results provided here support the brain-like network architecture hypothesis, which is that, the graph neural network can be improved by formulating its spatio-temporal processing module into a hierarchical structure.

\section{Pseudocode of NuerocircuitX}\label{sec:neux}
The detailed pseudocode of NuerocircuitX is provided in Algorithm~\ref{alg:NeuX}.
\begin{algorithm}[h]
	\caption{\label{alg:NeuX}BrainNetX}
	{\bf Input:} well-trained model $P(\cdot)$; cortical network parcellation atlas $F$; task-related brain state fMRI data $\boldsymbol{x}$\\
	{\bf Output:} importance score $\mathcal{I}$ of each neural circuit for all tasks
	\begin{algorithmic}[1]
	    \ForAll{$i=1, 2, \dots, n$}
	        \State Assign $\boldsymbol{x}_{i}$ into $Q$ cortical networks according to $F$ and obtain neural-circuit-wise input $F_{\boldsymbol{x}_{i}}=\{f_{\boldsymbol{x}_{i, 1}}, \dots,  f_{\boldsymbol{x}_{i, Q}}\}$
	    \EndFor
	    \ForAll{$i=1, 2, \dots, n$}
	        \ForAll{$q=1, 2, \dots, Q$}
	        \State Apply masking strategy to obtain $\bar{\phi}_{mask}(f_{\boldsymbol{x}_{i, q}})$
	        \State Apply keeping strategy to obtain $\bar{\phi}_{keep}(f_{\boldsymbol{x}_{i, q}})$
	        \State Calculate \emph{importance score} $\mathcal{I}(f_{\boldsymbol{x}_{i, q}})$
	        \EndFor
	    \EndFor
	    \State \Return $\mathcal{I}$
	\end{algorithmic} 
\end{algorithm}

\section{Additional Brain Annotation Results}~\label{sec:add_annotation}
In this section, we provide the brain annotation maps of ST-GCN and BrainNetX, as well as key brain regions annotated by Neurosynth.

For the annotation results derived from ST-GCN and BrainNetX (Fig.~\ref{fig:annotation_stgcn}), we noticed that for some tasks required low cognitive demand (\textit{e.g. motor tasks, language tasks}), ST-GCN performed roughly the same as STpGCN~\cite{ye2022spatial}. However, for higher-order cognitive tasks (\eg social cognition tasks), prior statistical studies on task-related brain states struggle to in line with the key brain areas found by ST-GCN.

To obtain the comparable annotation results for Neurosynth, we normalize the absolute value of all negative/positive (deactivated/activated) values for each brain region measured by Neurosynth. Therefore, we can determine the ``importance score'' of each brain region for each task evaluated by Neurosynth. The visualization of the annotation results for each category of tasks by Neurosynth are provided in Fig.~\ref{fig:neurosynth}.


\section{Setting about the Quantitative Explainability Experiments}~\label{sec:neurosynth}

In this section, we provide the details setting about the quantitative explainability experiments for Section~\uppercase\expandafter{\romannumeral5}-G in the main manuscript.

For state-of-the-art brain annotation method, Neurosynth, we transform its annotation results by normalizing the absolute value of all negative/positive (deactivated/activated) values for each brain region. For the key brain regions highlighted by each machine learning methods, we mask the exclusive regions (\textit{i.e.} set the node features of exclude regions as $0$) for each task to form as a distilled input to SVM-RBF to decode $8$ working memory task-related brain states (\textit{i.e.} each task will have at most $8\times k$ ROIs as input). To make the distilled input by Neurosynth comparable, for each $k$, we retain $8\times k$ regions and mask the features of rest regions.

\section{Details of the Annotation Results}~\label{sec:annotation_table}
In order to more intuitively show the key brain regions associated to the various task-related brain states found by different method, we provide the details of the brain regions in this section.

Table~\ref{tab:neurosynth} presents the annotated key brain regions of $7$ category of tasks using Neurosynth with MMP atlas which corresponds to Fig.~\ref{fig:neurosynth}. The core brain regions found by ST-GCN for $23$ task-related brain states are shown in Table~\ref{tab:stgcn} which corresponds to Fig.~\ref{fig:annotation_stgcn}. For Table~\ref{tab:annotation_stpgcn}, we provides the task-related brain regions investigated by STpGCN for $23$ task-related brain states, which are visualized in Fig. 5 of the main manuscript. Likewise, we offer the important brain regions of $23$ task-related brain states measured by MLP-Mixer (Table~\ref{tab:mlp_mixer}) which correspond to Fig. 6 of the main manuscript. For the stimuli classification experiments for examining the generalizability of STpGCN in the Section~\uppercase\expandafter{\romannumeral5}-H of the main manuscript, we give the details of core brain areas highlighted by STpGCN and BrainNetX in Table~\ref{tab:stpgcn_wm} which correspond to the Fig.~7 in the main manuscript.

\begin{table*}[htbp]
    \centering
    \caption{The annotated key brain regions using Neurosynth with MMP atlas.}\label{tab:neurosynth}
    \footnotesize
    \begin{tabular}{cc}
        \hline
        \textbf{Category} & \textbf{Annotated brain regions} \\
        \hline
        \makecell[c]{ Working\\Memory} & Inferior Parietal, Dorsolateral Prefrontal, Superior Parietal, Premotor, Inferior Frontal\\ \hline
        \makecell[c]{ Gambling} & Anterior Cingulate \& Medial Prefrontal, Orbital \& Polar Frontal, Dorsolateral Prefrontal\\\hline
        \makecell[c]{ Motor} & Paracentral Lobular \& Mid Cingulate, Somatosensory \& Motor, Premotor, Posterior Opercular\\
        \hline
        \makecell[c]{Language} & Auditory Association, Early Auditory\\
        \hline
        \makecell[c]{Social\\cognition} & \makecell[c]{Auditory Association, Inferior Parietal, Temporo Parieto Occipital Junction,\\ Anterior Cingulate \& Medial Prefrontal, Posterior Cingulate, Inferior Frontal}\\
        \hline
        \makecell[c]{Relational\\processing} & \makecell[c]{Dorsal Stream Visual, Inferior Parietal, Inferior Frontal,\\  Anterior Cingulate \& Medial Prefrontal, Orbital \& Polar Frontal, Dorsolateral Prefrontal}\\
        \hline
        \makecell[c]{Emotion\\processing} & \makecell[c]{Medial Temporal, Insular \& Frontal Opercular,\\ Anterior Cingulate \& Medial Prefrontal, Orbital \& Polar Frontal, Inferior Frontal}\\
        \hline
    \end{tabular}
\end{table*}

\begin{table*}[htbp]
    \centering
    \caption{The annotated key brain regions using ST-GCN on 15 time steps of fMRI with MMP atlas.}\label{tab:stgcn}
    \footnotesize
    \setlength{\tabcolsep}{1pt}
    \begin{tabular}{ccc}
        \hline
        \textbf{Category} & \textbf{Task} & \textbf{Annotated brain regions} \\
        \hline
        \multirow{8}{*}{\makecell[c]{ Working\\Memory}} & 0-back body   & \makecell[c]{Superior Parietal, Dorsal Stream Visual, MT+Complex \& Neighboring Visual Areas, Medial Temporal} \\ \cline{2-3} 
            & 0-back faces   & \makecell[c]{Inferior Frontal, Superior Parietal, Dorsolateral Prefrontal, Inferior Parietal, Dorsolateral Prefrontal} \\ \cline{2-3}
            & 0-back places   & \makecell[c]{Superior Parietal, Inferior Frontal, Anterior Cingulate \& Medial Prefrontal, Posterior Cingulate} \\ \cline{2-3} 
            & 0-back tools   & \makecell[c]{Medial Temporal, Posterio Cingulate} \\ \cline{2-3} 
            & 2-back body   & \makecell[c]{Inferior Frontal, Superior Parietal, Dorsolateral Prefrontal, Inferior Parietal} \\ \cline{2-3} 
            & 2-back faces   & \makecell[c]{Inferior Frontal, Superior Parietal, Dorsolateral Prefrontal, Inferior Parietal, Medial Temporal} \\ \cline{2-3}
            & 2-back places   & \makecell[c]{Orbital \& Polar Frontal,  Dorsolateral Prefrontal, Auditory Association, Lateral Temporal, Inferior Frontal} \\ \cline{2-3} 
            & 2-back tools   & \makecell[c]{Inferior Parietal, Inferior Frontal, Superior Parietal, Medial Temporal, Dorsolateral Prefrontal} \\ \hline
        \multirow{2}{*}{\makecell[c]{Gambling}} & Loss   & \makecell[c]{Early Auditory, Posterior Opercular, Insular \& Frontal Opercular, Medial Temporal} \\ \cline{2-3} 
            & Win   & \makecell[c]{Inferior Frontal, Superior Parietal, Dorsolateral Prefrontal, Inferior Parietal} \\ \hline
        \multirow{5}{*}{\makecell[c]{Motor}} & Tongue   & \makecell[c]{Insular \& Frontal Opercular, Posterior Opercular, Early Auditory} \\ \cline{2-3} 
            & Left foot   & \makecell[c]{Premotor, Somatosensory \& Motor, Paracentral Lobular \& Mid Cingulate} \\ \cline{2-3} 
            & Left hand   & \makecell[c]{Premotor, Somatosensory \& Motor, Paracentral Lobular \& Mid Cingulate} \\ \cline{2-3} 
            & Right hand   & \makecell[c]{Early Auditory, Posterior Opercular, Insular \& Frontal Opercular} \\ \cline{2-3} 
            & Right foot   & \makecell[c]{Early Auditory, Posterior Opercular, Insular \& Frontal Opercular} \\ \hline
        \multirow{2}{*}{Language} & Math   & \makecell[c]{Early Auditory, Posterior Opercular, Insular \& Frontal Opercular} \\ \cline{2-3} 
            & Story   & \makecell[c]{Early Auditory, Posterior Opercular, Insular \& Frontal Opercular} \\ \hline
        \multirow{2}{*}{\makecell[c]{Social\\cognition}} & \makecell[c]{Mental interaction}   & \makecell[c]{Early Auditory, Posterior Opercular Insular \& Frontal Opercular} \\ \cline{2-3} 
            & \makecell[c]{Random interaction }  & \makecell[c]{Superior Parietal, Dorsal Stream Visual, MT+Complex \& Neighboring Visual Areas, Medial Temporal} \\ \hline
        \multirow{2}{*}{\makecell[c]{Relational\\processing}} & Match   & \makecell[c]{Medial Temporal, Dorsal Stream Visual, Superior Parietal, MT+Complex \& Neighboring Visual Areas} \\ \cline{2-3} 
            & Relational   & \makecell[c]{Inferior Parietal, Inferior Frontal, Superior Parietal, Medial Temporal, Dorsolateral Prefrontal} \\ \hline
        \multirow{2}{*}{\makecell[c]{Emotion\\processing}} & \makecell[c]{Emotional face}   & \makecell[c]{Superior Parietal, Dorsal Stream Visual, MT+Complex \& Neighboring Visual Areas} \\ \cline{2-3} 
            & Shape   & \makecell[c]{Medial Temporal, Dorsal Stream Visual, Superior Parietal, MT+Complex \& Neighboring Visual Areas} \\ \hline
    \end{tabular}
\end{table*}

\begin{table*}[htbp]
    \centering
    \caption{The annotated key brain regions using STpGCN on 15 time steps of fMRI with MMP atlas.}\label{tab:annotation_stpgcn}
    \footnotesize
    \setlength{\tabcolsep}{1pt}
    \begin{tabular}{ccc}
        \hline
        \textbf{Category} & \textbf{Task} & \textbf{Annotated brain regions} \\
        \hline
        \multirow{8}{*}{\makecell[c]{Working\\Memory}} & 0-back body   & \makecell[c]{Superior Parietal, Dorsal Stream Visual, MT+Complex \& Neighboring Visual Areas} \\ \cline{2-3} 
            & 0-back faces   & \makecell[c]{Inferior Frontal, Superior Parietal, Dorsolateral Prefrontal, Inferior Parietal} \\ \cline{2-3}
            & 0-back places   & \makecell[c]{Insular \& Frontal Opercular, Inferior Frontal, Anterior Cingulate \& Medial Prefrontal, Dorsolateral Prefrontal} \\ \cline{2-3} 
            & 0-back tools   & \makecell[c]{Medial Temporal, Posterio Cingulate, Temporo Parieto Occipital Junction} \\ \cline{2-3} 
            & 2-back body   & \makecell[c]{Insular \& Frontal Opercular, Posterior Opercular, Early Auditory} \\ \cline{2-3} 
            & 2-back faces   & \makecell[c]{Inferior Frontal, Superior Parietal, Dorsolateral Prefrontal, Inferior Parietal} \\ \cline{2-3}
            & 2-back places   & \makecell[c]{Orbital \& Polar Frontal,  Dorsolateral Prefrontal, Superior Parietal, Lateral Temporal, Inferior Frontal, Inferior Parietal} \\ \cline{2-3} 
            & 2-back tools   & \makecell[c]{Posterior Cingulate, Anterior Cingulate \& Medial Prefrontal} \\ \hline
        \multirow{2}{*}{\makecell[c]{Gambling}} & Loss   & \makecell[c]{Inferior Frontal, Superior Parietal, Dorsolateral Prefrontal, Inferior Parietal} \\ \cline{2-3} 
            & Win   & \makecell[c]{Inferior Frontal, Superior Parietal, Dorsolateral Prefrontal, Inferior Parietal} \\ \hline
        \multirow{5}{*}{\makecell[c]{Motor}} & Tongue   & \makecell[c]{Insular \& Frontal Opercular, Posterior Opercular, Early Auditory} \\ \cline{2-3} 
            & Left foot   & \makecell[c]{Premotor, Somatosensory \& Motor, Paracentral Lobular \& Mid Cingulate} \\ \cline{2-3} 
            & Left hand   & \makecell[c]{Premotor, Somatosensory \& Motor, Paracentral Lobular \& Mid Cingulate} \\ \cline{2-3} 
            & Right hand   & \makecell[c]{Premotor, Somatosensory \& Motor, Paracentral Lobular \& Mid Cingulate} \\ \cline{2-3} 
            & Right foot   & \makecell[c]{Premotor, Somatosensory \& Motor, Posterior Opercular, Early Auditory} \\ \hline
        \multirow{2}{*}{Language} & Math   & \makecell[c]{Insular \& Frontal Opercular, Posterior Opercular, Early Auditory} \\ \cline{2-3} 
            & Story   & \makecell[c]{Insular \& Frontal Opercular, Posterior Opercular, Early Auditory} \\ \hline
        \multirow{2}{*}{\makecell[c]{Social\\cognition}} & \makecell[c]{Mental interaction}   & \makecell[c]{Superior Parietal, Dorsal Stream Visual, MT+Complex Neighboring Visual Areas} \\ \cline{2-3} 
            & \makecell[c]{Random interaction }  & \makecell[c]{Superior Parietal, Dorsal Stream Visual, MT+Complex \& Neighboring Visual Areas} \\ \hline
        \multirow{2}{*}{\makecell[c]{Relational\\processing}} & Match   & \makecell[c]{Inferior Frontal, Superior Parietal, Dorsolateral Prefrontal, Inferior Parietal} \\ \cline{2-3} 
            & Relational   & \makecell[c]{Insular \& Frontal Opercular, Posterior Opercular, Early Auditory} \\ \hline
        \multirow{2}{*}{\makecell[c]{Emotion\\processing}} & \makecell[c]{Emotional face}   & \makecell[c]{Superior Parietal, Dorsal Stream Visual, MT+Complex \& Neighboring Visual Areas} \\ \cline{2-3} 
            & Shape   & \makecell[c]{Posterior Opercular, Superior Parietal, Early Auditory, Inferior Frontal, Inferior Parietal} \\ \hline
    \end{tabular}
\end{table*}

\begin{table*}[htbp]
    \centering
    \caption{The annotated key brain regions using MLP-Mixer on 15 time steps of fMRI with MMP atlas.}\label{tab:mlp_mixer}
    \footnotesize
    \setlength{\tabcolsep}{1pt}
    \begin{tabular}{ccc}
        \hline
        \textbf{Category} & \textbf{Task} & \textbf{Annotated brain regions} \\
        \hline
        \multirow{8}{*}{\makecell[c]{ Working\\Memory}} & 0-back body   & \makecell[c]{Inferior Parietal, Dorsal Stream Visual, MT+Complex \& Neighboring Visual Areas, Superior Parietal} \\ \cline{2-3} 
            & 0-back faces   & \makecell[c]{Ventral Stream Visual, MT+Complex \& Neighboring Visual Areas, Early Visual} \\ \cline{2-3}
            & 0-back places   & \makecell[c]{Ventral Stream Visual, MT+Complex \& Neighboring Visual Areas, Early Visual} \\ \cline{2-3} 
            & 0-back tools   & \makecell[c]{Ventral Stream Visual, MT+Complex \& Neighboring Visual Areas, Early Visual} \\ \cline{2-3} 
            & 2-back body   & \makecell[c]{Ventral Stream Visual, MT+Complex \& Neighboring Visual Areas, Early Visual} \\ \cline{2-3} 
            & 2-back faces   & \makecell[c]{Ventral Stream Visual, MT+Complex \& Neighboring Visual Areas, Early Visual} \\ \cline{2-3}
            & 2-back places   & \makecell[c]{MT+Complex \& Neighboring Visual Areas, Dorsal Stream Visual, Superior Parietal} \\ \cline{2-3} 
            & 2-back tools   & \makecell[c]{Ventral Stream Visual, MT+Complex \& Neighboring Visual Areas, Early Visual} \\ \hline
        \multirow{2}{*}{\makecell[c]{Gambling}} & Loss   & \makecell[c]{Ventral Stream Visual, MT+Complex \& Neighboring Visual Areas, Early Visual} \\ \cline{2-3} 
            & Win   & \makecell[c]{Ventral Stream Visual, MT+Complex \& Neighboring Visual Areas, Early Visual} \\ \hline
        \multirow{5}{*}{\makecell[c]{Motor}} & Tongue   & \makecell[c]{Insular \& Frontal Opercular, Early Auditory, Posterior Opercular} \\ \cline{2-3} 
            & Left foot   & \makecell[c]{Premotor, Somatosensory \& Motor, Paracentral Lobular \& Mid Cingulate} \\ \cline{2-3} 
            & Left hand   & \makecell[c]{Premotor, Somatosensory \& Motor, Paracentral Lobular \& Mid Cingulate} \\ \cline{2-3} 
            & Right hand   & \makecell[c]{Ventral Stream Visual, MT+Complex \& Neighboring Visual Areas, Early Visual} \\ \cline{2-3} 
            & Right foot   & \makecell[c]{Insular \& Frontal Opercular, Early Auditory, Posterior Opercular} \\ \hline
        \multirow{2}{*}{Language} & Math   & \makecell[c]{Insular \& Frontal Opercular, Early Auditory, Posterior Opercular} \\ \cline{2-3} 
            & Story   & \makecell[c]{Insular \& Frontal Opercular, Early Auditory, Posterior Opercular} \\ \hline
        \multirow{2}{*}{\makecell[c]{Social\\cognition}} & \makecell[c]{Mental interaction}   & \makecell[c]{MT+Complex \& Neighboring Visual Areas, Inferior Parietal, Dorsal Stream Visual, Superior Parietal} \\ \cline{2-3} 
            & \makecell[c]{Random interaction }  & \makecell[c]{MT+Complex \& Neighboring Visual Areas, Dorsal Stream Visual, Superior Parietal} \\ \hline
        \multirow{2}{*}{\makecell[c]{Relational\\processing}} & Match   & \makecell[c]{Ventral Stream Visual, MT+Complex \& Neighboring Visual Areas, Early Visual} \\ \cline{2-3} 
            & Relational   & \makecell[c]{Ventral Stream Visual, MT+Complex \& Neighboring Visual Areas, Early Visual} \\ \hline
        \multirow{2}{*}{\makecell[c]{Emotion\\processing}} & \makecell[c]{Emotional face}   & \makecell[c]{Ventral Stream Visual, MT+Complex \& Neighboring Visual Areas, Early Visual} \\ \cline{2-3} 
            & Shape   & \makecell[c]{Ventral Stream Visual, MT+Complex \& Neighboring Visual Areas, Early Visual} \\ \hline
    \end{tabular}
\end{table*}

\begin{table*}[htbp]
    \centering
    \caption{The annotated key brain regions of different working memory task-related brain states with the same stimuli by STpGCN and BrainNetX given 15 time steps of fMRI with MMP atlas across 10-fold cross validation.}\label{tab:stpgcn_wm}
    \footnotesize
    \begin{tabular}{cc}
        \hline
        \textbf{\makecell[c]{Visual\\stimuli}} & \textbf{Annotated brain regions} \\
        \hline
        \makecell[c]{body} & Inferior Parietal, Dorsolateral Prefrontal, Superior Parietal, Premotor, Inferior Frontal\\ \hline
        \makecell[c]{faces} & Inferior Parietal, Dorsolateral Prefrontal, Superior Parietal, Premotor, Inferior Frontal\\
        \hline
        \makecell[c]{places} & Superior Parietal, Posterior Cingulate\\
        \hline
        \makecell[c]{tools} & Early Auditory, Insular \& Frontal Opercular, Posterior Opercular\\
        \hline
    \end{tabular}
\end{table*}


\begin{table*}[htbp]
	\centering
	\begin{threeparttable}[b]
	\caption{\label{tab:top_k_roi_svm_comparison}
	Using the union of top K ROIs of different working memory tasks obtained by distinct models with NeurocircuitX as input to train SVM-RBF with 64 max iteration and one-versus-rest decision policy.
	}
    \setlength{\tabcolsep}{1.6pt}
	\scriptsize
	\begin{tabular}{c|cccc|cccc|cccc}
	\toprule
		\multirow{1}{*}{\diagbox{K}{Models}} & 
		\multicolumn{4}{c|}{MLP-Mixer+NeurocircuitX} &
		\multicolumn{4}{c|}{ST-GCN+NeurocircuitX} &
		\multicolumn{4}{c}{STpGCN+NeurocircuitX}\\
		& ACC$\uparrow$ & Macro Pre$\uparrow$ & Macro R$\uparrow$ & Macro F1$\uparrow$ & ACC$\uparrow$ & Macro Pre$\uparrow$ & Macro R$\uparrow$ & Macro F1$\uparrow$ & ACC$\uparrow$    & Macro Pre$\uparrow$ & Macro R$\uparrow$ & Macro F1$\uparrow$\\
		\midrule
		1 &  32.3$\pm$1.6 & 34.2$\pm$1.5 & 32.3$\pm$1.6 & 30.8$\pm$2.2 & 42.8$\pm$3.9 & 45.6$\pm$4.6 & 42.8$\pm$3.9 & 41.9$\pm$4.3 & 46.8$\pm$2.4  & 48.6$\pm$2.8  & 46.8$\pm$2.4  & 46.4$\pm$2.5\\
		2 &  34.7$\pm$2.2 & 36.5$\pm$2.7 & 34.7$\pm$2.2 & 33.8$\pm$3.0 & 43.9$\pm$2.1 & 45.7$\pm$2.5 & 43.9$\pm$2.1 & 43.5$\pm$2.2 & 51.6$\pm$2.7  & 54.1$\pm$2.4  & 51.6$\pm$2.7  & 51.4$\pm$3.0\\
		3 &  37.6$\pm$1.9 & 39.6$\pm$2.1 & 37.6$\pm$1.9 & 36.7$\pm$2.3 & 52.9$\pm$2.0 & 55.2$\pm$2.2 & 52.9$\pm$2.0 & 52.7$\pm$2.5 & 59.3$\pm$3.2  & 61.2$\pm$3.4  & 59.3$\pm$3.2  & 59.4$\pm$3.4\\
		4 &  38.8$\pm$1.5 & 41.8$\pm$2.7 & 38.8$\pm$1.5 & 38.1$\pm$1.5 & 55.9$\pm$2.9 & 57.5$\pm$3.0 & 55.9$\pm$2.9 & 55.8$\pm$2.8 & 61.1$\pm$2.0  & 63.2$\pm$1.8  & 61.1$\pm$2.0  & 61.3$\pm$2.1\\
		5 &  39.6$\pm$2.4 & 42.9$\pm$3.3 & 39.6$\pm$2.4 & 38.9$\pm$2.9 & 57.1$\pm$2.9 & 59.0$\pm$3.5 & 57.1$\pm$2.9 & 56.8$\pm$3.3 & 61.5$\pm$2.3  & 63.3$\pm$2.4  & 61.5$\pm$2.3  & 61.6$\pm$2.4\\
		6 &  41.6$\pm$2.3 & 43.9$\pm$3.2 & 41.6$\pm$2.3 & 41.0$\pm$2.8 & 61.6$\pm$1.9 & 62.8$\pm$1.8 & 61.9$\pm$1.9 & 61.5$\pm$1.8 & 63.8$\pm$2.0  & 65.3$\pm$2.4  & 63.7$\pm$2.0  & 64.0$\pm$2.1\\
		7 &  41.2$\pm$2.7 & 43.1$\pm$3.0 & 41.2$\pm$2.7 & 40.6$\pm$3.2 & 62.2$\pm$2.3 & 64.3$\pm$1.8 & 62.2$\pm$2.2 & 62.2$\pm$2.3 & 64.8$\pm$3.3  & 65.9$\pm$3.2  & 64.8$\pm$3.3  & 64.8$\pm$3.4\\
		8 &  42.8$\pm$1.7 & 45.3$\pm$2.1 & 42.6$\pm$1.7 & 42.2$\pm$1.9 & 63.0$\pm$2.6 & 65.1$\pm$2.6 & 63.0$\pm$2.6 & 63.2$\pm$2.6 & 64.1$\pm$2.4  & 65.9$\pm$1.9  & 64.1$\pm$2.4  & 64.2$\pm$2.4\\
		9 &  42.8$\pm$1.6 & 45.3$\pm$2.1 & 42.8$\pm$1.7 & 42.2$\pm$1.8 & 61.7$\pm$3.2 & 63.9$\pm$2.7 & 61.7$\pm$3.2 & 61.7$\pm$3.2 & 64.9$\pm$1.7  & 67.0$\pm$2.0  & 65.0$\pm$1.7  & 65.0$\pm$1.8\\
		10 &  43.2$\pm$3.3 &45.7$\pm$3.5 & 43.2$\pm$3.3 & 42.8$\pm$3.4 & 64.3$\pm$2.7 & 65.8$\pm$2.6 & 64.3$\pm$2.7 & 64.5$\pm$2.7 & 64.7$\pm$1.7  & 67.2$\pm$2.2  & 64.7$\pm$1.7  & 64.8$\pm$1.9\\
		\bottomrule
	\end{tabular}
	\end{threeparttable}
\end{table*}

\begin{table*}[htbp]
    \centering
    \caption{Comparison of the annotation similarity under different metrics  between different models together with NeurocircuitX on 15 time steps of fMRI with MMP atlas and the Neurosynth. The best results are highlighted with boldface. }\label{tab:result_annotation}
    \setlength{\tabcolsep}{20pt}
    \begin{tabular}{ccccc}
        \hline
        Methods         & PCC ($\%$) $\uparrow$    & Cosine dist ($\%$) $\downarrow$ & MSE$\downarrow$\\ \hline
        \underline{\textbf{Baselines:}} &&&&\\
        GCN           & 27.1$\pm$30.0 & 58.1$\pm$23.3     &   93.3$\pm$33.7\\
        GAT            & 31.3$\pm$26.8  & 56.3$\pm$22.1     &  82.0$\pm$29.2\\
        ST-GCN         & 31.7$\pm$28.2  & 56.1$\pm$23.4     &  \textbf{72.5$\pm$24.6}\\ \hline
        \underline{\textbf{Proposed:}} &&&&\\
        \textbf{STpGCN} & \textbf{33.0$\pm$27.0}  & \textbf{54.9$\pm$22.6}     & 79.3$\pm$25.2    &\\ \hline
    \end{tabular}
\end{table*}

\begin{table*}[htbp]
    \centering
    \begin{threeparttable}[b]
    \caption{Comparison of the 23 task-related brain states decoding results ($\%$) of our model and some baseline models on 15 time steps of fMRI with \textbf{AAL} atlas across 10-fold cross-validation. The best results are highlighted with boldface. }\label{tab:classification_result_aal}
    \setlength{\tabcolsep}{20pt}
    \begin{tabular}{ccccc}
        \hline
        Methods         & ACC$\uparrow$    & Macro Pre$\uparrow$ & Macro R$\uparrow$ & Macro F1$\uparrow$ \\ \hline
        \underline{\textbf{Baselines:}} &&&&\\
        SVM-RBF  & 57.9$\pm$1.8  & 63.7$\pm$1.7     &   57.9$\pm$1.8  & 57.8$\pm$1.9   \\
        MLP-Mixer  & 81.7$\pm$1.4  & 82.4$\pm$1.2     &   81.7$\pm$1.4  & 81.6$\pm$1.4   \\
        GCN           & 37.4$\pm$1.4  & 36.2$\pm$1.3     &   37.4$\pm$1.3  & 36.3$\pm$1.3   \\
        GIN  & 64.1$\pm$2.7  & 64.6$\pm$3.0     &   64.2$\pm$2.7  & 63.7$\pm$2.8   \\
        GAT            & 55.0$\pm$3.6  & 55.0$\pm$3.5     &  55.0$\pm$3.6   &   53.8$\pm$4.4 \\
        ST-GCN         & 74.9$\pm$1.8  & 75.1$\pm$1.7     &  75.0$\pm$1.8   &   74.8$\pm$1.8 \\ \hline
        \underline{\textbf{Proposed:}} &&&&\\
        STpGCN-$\alpha$       & 81.8$\pm$2.2  & 82.0$\pm$2.1     &  81.8$\pm$2.1   &   81.8$\pm$2.2 \\
        STpGCN-$\beta$       & 81.4$\pm$1.5  & 81.5$\pm$1.5     &  81.3$\pm$1.5   &   81.3$\pm$1.5 \\ 
        STpGCN-$\gamma$       & 80.3$\pm$2.6  & 80.4$\pm$2.6     &  80.3$\pm$2.6   &   80.2$\pm$2.6 \\
        \textbf{STpGCN} & \textbf{82.3$\pm$1.6}  & \textbf{82.5$\pm$1.6}     & \textbf{82.3$\pm$1.6}    & \textbf{82.3$\pm$1.6}   \\ \hline
    \end{tabular}
    \begin{tablenotes}[flushleft]
     \item Note: STpGCN-$\alpha$, STpGCN-$\beta$ and STpGCN-$\gamma$ denote STpGCN without mid spatial-temporal pathway, STpGCN without top spatial-temporal pathway and STpGCN without bottom-up pathway, respectively.
   \end{tablenotes}
   \end{threeparttable}
\end{table*}

\begin{table*}[htbp]
    \centering
    \begin{threeparttable}[b]
    \caption{Comparison of the 23 task-related brain states decoding results ($\%$) of our model and some baseline models on 15 time steps of fMRI with \textbf{MMP} atlas across 10-fold cross-validation. The best results are highlighted with boldface. }\label{tab:classification_result_mmp}
    \setlength{\tabcolsep}{20pt}
    \begin{tabular}{ccccc}
        \hline
        Methods         & ACC$\uparrow$    & Macro Pre$\uparrow$ & Macro R$\uparrow$ & Macro F1$\uparrow$ \\ \hline
        (a) \underline{\textbf{Baselines:}} &&&&\\
        SVM-RBF  & 79.9$\pm$1.6  & 81.8$\pm$1.3     &   80.0$\pm$1.6  & 80.0$\pm$1.6   \\
        MLP-Mixer  & 89.8$\pm$1.0  & 89.9$\pm$0.9     &   89.8$\pm$0.9  & 89.8$\pm$0.9   \\
        GCN           & 61.4$\pm$1.6  & 60.8$\pm$1.5     &   61.4$\pm$1.6  & 60.7$\pm$1.5   \\
        GIN  & 80.1$\pm$2.1  & 80.6$\pm$2.1     &   80.1$\pm$2.1  & 80.1$\pm$2.1   \\
        GAT            & 82.2$\pm$1.2  & 82.3$\pm$1.2     &  82.3$\pm$1.3   &   82.2$\pm$1.3 \\
        ST-GCN         & 90.6$\pm$0.6  & 90.7$\pm$0.6     &  90.6$\pm$0.6   &   90.6$\pm$0.6 \\ \hline
        \underline{\textbf{Proposed:}} &&&&\\
        (b) STpGCN-$\alpha$       & 91.0$\pm$0.6  & 91.1$\pm$0.6     &  91.0$\pm$0.6   &   91.0$\pm$0.6 \\
        (c) STpGCN-$\beta$       & 91.2$\pm$0.8  & 91.3$\pm$0.8     &  91.3$\pm$0.8   &   91.0$\pm$0.8 \\ 
        (d) STpGCN-$\gamma$       & 91.2$\pm$0.8  & 91.3$\pm$0.8     &  91.3$\pm$0.8   &   91.0$\pm$0.8 \\ 
        (e) \textbf{STpGCN} & \textbf{92.4$\pm$0.8}  & \textbf{92.5$\pm$0.7}     & \textbf{92.4$\pm$0.8}    & \textbf{92.4$\pm$0.7}   \\ \hline
    \end{tabular}
    \begin{tablenotes}[flushleft]
     \item Note: STpGCN-$\alpha$, STpGCN-$\beta$ and STpGCN-$\gamma$ denote STpGCN without mid spatial-temporal pathway, STpGCN without top spatial-temporal pathway and STpGCN without bottom-up pathway, respectively.
   \end{tablenotes}
   \end{threeparttable}
\end{table*}

\end{document}